\documentclass[aps,prfluids,onecolumn,10pt,amsmath,amssymb,floatfix,superscriptaddress,longbibliography]{revtex4-2}
\usepackage[T1]{fontenc}
\usepackage[utf8]{inputenc}
\usepackage{graphicx,bm,mathptmx,helvet}
\usepackage{xcolor}
\usepackage{float}
\usepackage{array}
\usepackage[hidelinks]{hyperref}

\newcommand{\externalrefnumber}[2]{\expandafter\gdef\csname r@#1\endcsname{{#2}{}{}{}{}}}
\AtBeginDocument{
  \externalrefnumber{S-fig:s_advective}{S8}
  \externalrefnumber{S-fig:s_device}{S7}
  \externalrefnumber{S-fig:s_exchange}{S4}
  \externalrefnumber{S-fig:s_limits}{S6}
  \externalrefnumber{S-fig:s_pressure}{S1}
  \externalrefnumber{S-sec:darcy_solver}{S2}
  \externalrefnumber{S-sec:derivation}{S1}
  \externalrefnumber{S-sec:numerics}{S3}
  \externalrefnumber{S-tab:verification}{S3}
  \externalrefnumber{S-tab:balance}{S6}
  \externalrefnumber{S-tab:g1refine}{S5}
  \externalrefnumber{S-fig:s_r1t}{S9}
  \externalrefnumber{S-sec:r1t}{S7}
  \externalrefnumber{S-sec:verification}{S5}
  \externalrefnumber{S-tab:recon}{S7}
  \externalrefnumber{S-tab:metric}{S2}
  \externalrefnumber{S-fig:s_tracer}{S2}
  \externalrefnumber{S-sec:plug_transport}{S2\,B}
  \externalrefnumber{S-tab:theoryranges}{S4}
}
\graphicspath{{figures/}}
\newcommand{\tid}{t_{\mathrm{ideal}}}
\newcommand{\te}{\tau_e}
\newcommand{\Pe}{\mathrm{Pe}}

\begin{document}

\title{Pressure demand and solute replacement in a collagen-plugged bioreactor}

\author{Mohammad Mehedi Hasan Akash}
\email[Contact author: ]{makash@eng.famu.fsu.edu}
\affiliation{Department of Mechanical and Aerospace Engineering, FAMU--FSU College of Engineering, Tallahassee, Florida, USA}
\affiliation{Department of Mechanical Engineering, South Dakota State University, Brookings, South Dakota, USA}
\author{Mohammad Yeasin}
\affiliation{Department of Mechanical Engineering, South Dakota State University, Brookings, South Dakota, USA}
\author{Karl Van Horsen}
\affiliation{Department of Pharmaceutical Sciences, North Dakota State University, Fargo, North Dakota, USA}
\author{Mithul Sathish}
\affiliation{Department of Pharmaceutical Sciences, North Dakota State University, Fargo, North Dakota, USA}
\author{Susan Immanuel}
\affiliation{Department of Pharmaceutical Sciences, College of Pharmacy and Allied Health Professions, South Dakota State University, Brookings, South Dakota, USA}
\author{Sanku Mallik}
\affiliation{Department of Pharmaceutical Sciences, College of Pharmacy and Allied Health Professions, South Dakota State University, Brookings, South Dakota, USA}
\author{Saikat Basu}
\email[Contact author: ]{Saikat.Basu@sdstate.edu}
\affiliation{Department of Mechanical Engineering, South Dakota State University, Brookings, South Dakota, USA}
\date{25 November 2026}

\begin{abstract}
In a perfusion bioreactor, fresh culture medium enters through a small opening,
flows through a collagen plug, and leaves through a second opening. How much
pressure is needed to sustain that flow? When fresh medium reaches the outlet,
has it also replaced the old medium inside the pores? To answer these questions, we simulated flow and medium replacement in pores
traced from a collagen image. The flow simulations and curved-channel theory
show how narrow, winding passages concentrate resistance. We used that
resistance to estimate the plug permeability and found that the modeled plug
accounts for most of the pressure drop across the chamber. Feeding through
small openings adds to that demand because the flow must spread through the
plug and converge again at the outlet. At the same flow rate, this requires
about seven times the pressure needed with uniform feeding across the plug
faces. Pore shape also affects how closely the outlet signal tracks replacement in
the interior. In most runs, the interior reaches 99\% replacement
before the outlet does. A slowly exchanging side pocket can reverse that
order by retaining old medium while fresh medium passes through the main
channel. In the strongest case, the interior takes nearly three times as
long as the outlet to reach the same replacement threshold.
\end{abstract}
\maketitle

\section{Introduction}\label{sec:introduction}
Tumor cells respond to both the structure of the surrounding collagen
matrix and the interstitial flow through it. Collagen alignment and
crosslinking can promote invasion~\cite{provenzano2006,provenzano2008,levental2009}. Interstitial
flow can guide migration through competing mechanical and biochemical
cues~\cite{polacheck2011,shieh2011}, while matrix remodeling continuously
changes the pore space through which fluid and cells move~\cite{winkler2020}.
In vitro tumor models therefore need control of the hydraulic environment
as well as of the biological preparation~\cite{malandrino2018,swartz2012}.
Microfluidic and organ-on-chip platforms provide such control at small
scales~\cite{whitesides2006,bhatia2014,huh2010}, and perfusion bioreactors
extend it to millimeter-scale collagen plugs that can hold spheroids.
Perfused three-dimensional breast-cancer models of this
kind are being developed for preclinical testing~\cite{marshall2017,goliwas2017},
and the bioreactor that motivates the present study was built to support
studies of breast cancer metastasis in collagen.

In a perfused plug, however, the pump setting is only an indirect
control. It fixes neither the pressure the plug requires nor the time at
which the fluid inside the pores is actually replaced. Scaffold
architecture affects the biological response~\cite{obrien2005}, and
perfusion design depends on the local flow rather than the external flow
rate alone~\cite{zhao2018}.
Pore-resolved scaffold simulations have supplied permeabilities for
device-scale Darcy models~\cite{zhao2019multiscale}. More recent CFD-guided
bioreactor studies have examined how geometry and pressure control
interstitial-flow distributions during scale-up~\cite{vatani2026}.

Two features of a collagen pore space make the
gap between setting and outcome large. Narrow connected throats
concentrate the viscous resistance, so a small change in throat width
changes the pressure demand strongly. Regions that exchange slowly with
those throats can hold old fluid after the outlet concentration has
nearly reached its inlet value. Our earlier reduced-order models examined
perfusion and solute transport in biomimetic dense tumor
matrices~\cite{akash2023ecmf,akash2026fphar}.
Here we connect the resistance of individual pores
to the pressure required to perfuse the plug through the device ports.

Classical lubrication theory~\cite{reynolds1886} and
slow-variation expansions~\cite{vandyke1987} relate channel shape to
resistance, with extensions to periodic pores~\cite{kitanidis1997},
higher-order Cartesian channels~\cite{tavakol2017,banerjee2024}, curved
substrates~\cite{roy2002} and rough curved channels~\cite{okechi2020}.
Comparisons with rational-approximation Stokes solutions have quantified
the error of higher-order lubrication predictions in smoothly constricted
channels~\cite{xue2024}. For wavy channels containing a porous medium,
Dewangan~\cite{dewangan2025} compared lubrication and spectral predictions
within a Darcy--Brinkman model.
Upscaled to a plug, the pore resistance gives a permeability, and Darcy's
law gives the pressure~\cite{bear1972,carman1997}. Pores traced from an
image, however, wind strongly and have walls that are neither parallel nor
centered on any fixed axis. Which reference line to follow, how to treat
the offset of the walls from it, and over which interval to compare with
simulations then become practical questions that the existing theories do
not settle.

For solute replacement, residence-time and mobile--immobile exchange
theory describe how slowly exchanging regions broaden an outlet
response~\cite{danckwerts1953,coats1964}.
For shallow microchannels, Ajdari, Bontoux and Stone~\cite{ajdari2006}
identified different geometric controls for hydraulic resistance and
dispersion: the former depends on the thin gap, whereas the latter can be
controlled by the larger transverse width.
This distinction motivates examining how the
pores are refreshed as well as how they resist flow. The outlet concentration
is weighted by the fluid leaving the pore, and cells experience the local
concentration inside it. Whether the outlet signal is early or late relative to the
interior therefore depends on the pore geometry and on how fast solute
crosses streamlines; here we examine it in pores traced from a collagen
image.

The present study addresses both gaps for one chamber of a perfusion
bioreactor,
with a narrow inlet conduit, a collagen-holding chamber and an outlet
port. Operators reported stalled or reversed flow at high pump settings, and
dye downstream of the plug in a test at a 5~mL/min pump setting, which motivates two questions:
\begin{itemize}
\item[] \hypertarget{intro-q1}{}\textbf{(Q1)} \emph{What sets the pressure needed to perfuse the collagen plug?}
\item[] \hypertarget{intro-q2}{}\textbf{(Q2)} \emph{When the outlet shows fresh medium, has the fluid inside the
pores been replaced?}
\end{itemize}
The pressure calculation begins with
\emph{open-device simulations}, which establish the pressure loss through the
chamber and conduits with fluid alone. The
\emph{pore simulations} in five two-dimensional channels traced from one
collagen image supply the
resistance used to estimate plug permeability (Sec.~\ref{sec:perm}). We use
\emph{a curved-channel lubrication theory} to predict the
pore resistance from geometry and identify which features control it
(Sec.~\ref{sec:curved}). We then use
\emph{a Darcy model of the plug} to combine permeability with feeding geometry
and obtain the pressure demand (Sec.~\ref{sec:darcy}). We check that demand
against a three-dimensional porous-plug simulation.

\begin{figure}[!tbp]
\centering\includegraphics[width=\textwidth]{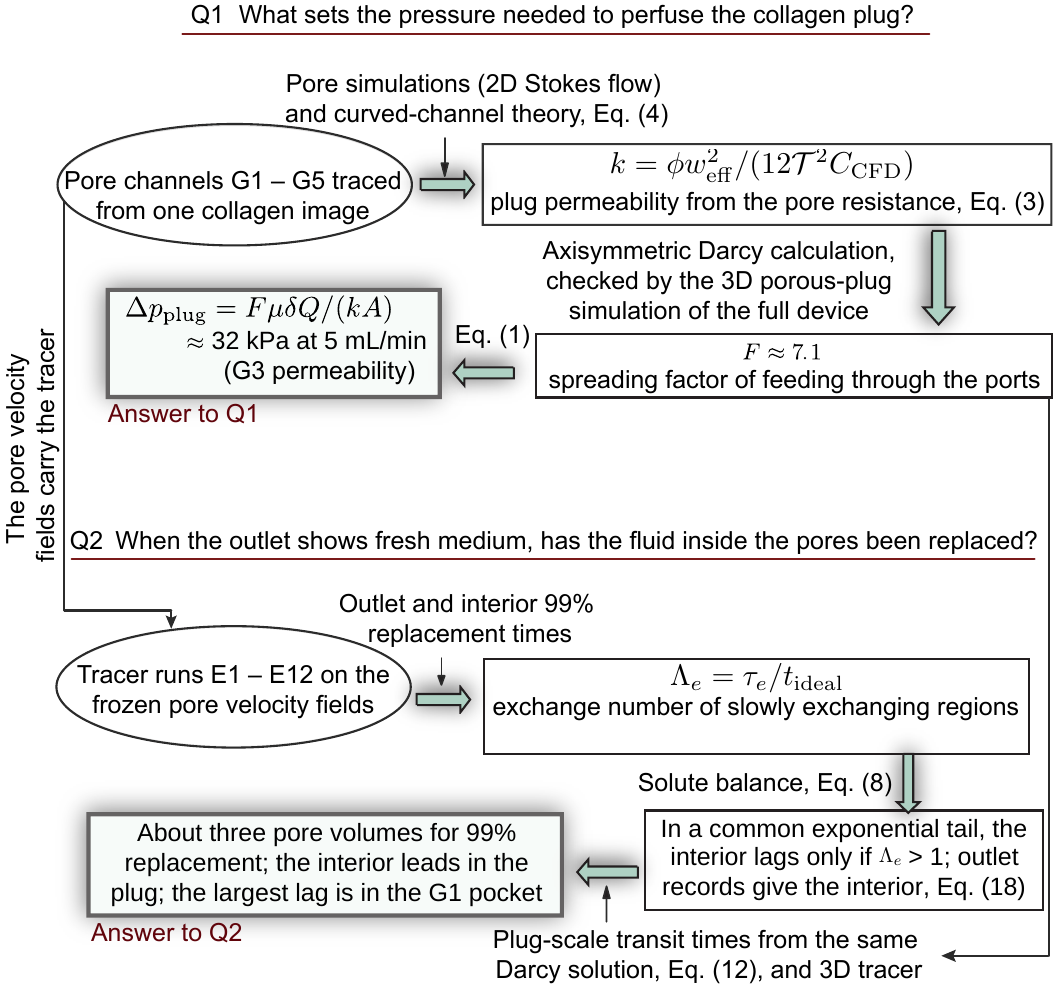}
\caption{\label{fig:roadmap}Two chains of calculations answer the two
questions. Top (Q1): the pore resistance of the traced channels gives the
plug permeability, the axisymmetric Darcy calculation gives the spreading
factor of port feeding, and together they give the pressure demand of the
plug. Bottom (Q2): the tracer runs give the exchange number of slowly
exchanging regions, the solute balance turns it into a criterion for when
the interior lags the outlet, and the plug-scale calculations carry it to
the device. Broad arrows mark the main steps and thin arrows the
supporting simulations; the two thin links show what the chains share:
the pore velocity fields carry the tracer, and the Darcy solution that
gives the spreading factor also gives the plug-scale transit times.
Flow rates refer to the flow through the modeled chamber-filling plug.}
\end{figure}

The
\emph{tracer simulations} reuse the pore velocity fields to compare outlet
and internal replacement. The solute balance relates the two histories
and gives a criterion for lag when the replacement threshold falls in a
common exponential tail (Sec.~\ref{sec:replacement}). At the plug scale,
the Darcy flow field also gives the advective transit times.
The experimental apparatus supplies the geometry, operating range and
biological context (Sec.~\ref{sec:experiment});
no measured pressure or concentration is used for calibration or
validation. Figure~\ref{fig:roadmap} shows how these calculations
lead to the two answers.

\section{Theory}\label{sec:theory}
The two questions lead to two strands of theory. The pressure needed to
perfuse the plug depends on how fluid is fed and on the plug
permeability, which in turn is set by the resistance of individual pores
(Secs.~\ref{sec:darcy}--\ref{sec:curved}). Whether perfusion replaces the
interior fluid depends instead on transit and exchange times within those
pores (Sec.~\ref{sec:replacement}). Each element of the theory is fed or
tested by a specific simulation, as stated where it is introduced.

\subsection{Pressure demand of the plug}\label{sec:darcy}
For a homogeneous plug of permeability $k$,\ radius $R$, face area $A=\pi R^2$ and
thickness $\delta$, perfused with fluid of viscosity $\mu$ at flow rate
$Q$, Darcy's law~\cite{bear1972} with uniform feeding over the whole face
gives the resistance $\mu\delta/(kA)$. When fluid enters and leaves
through smaller ports, it must spread through the plug, and the
resistance increases by a spreading factor $F\geq1$. The pressure demand
and its ratio to the pressure $\Delta P_c$ that the circuit can supply
across the plug are
\begin{equation}
\Delta p_{\rm plug}=\frac{F\mu\delta Q}{kA},\qquad
\Pi_P=\frac{\Delta p_{\rm plug}}{\Delta P_c}.
\label{eq:capacity}
\end{equation}
A flow is pressure-feasible when $\Pi_P\leq1$, so the largest
sustainable flow is $Q_{\max}=Q/\Pi_P$. For a cylinder fed through two
coaxial disks of radii $a_1,a_2$ with all other boundaries sealed, the
pressure satisfies Laplace's equation, as in electrical and thermal
constriction problems~\cite{holm1967,cooper1969}; we obtain $F$ by solving
it numerically in an axisymmetric Darcy calculation and test it with a
three-dimensional porous-plug simulation of the full device, which checks
the feeding reduction at the prescribed $k$ but not the permeability
closure.

For constant isotropic permeability and viscosity, Darcy's law
$\mathbf u=-(k/\mu)\nabla p$ and incompressibility give the axisymmetric
governing equation
\begin{equation}
\frac{1}{r}\frac{\partial}{\partial r}
\left(r\frac{\partial p}{\partial r}\right)
+\frac{\partial^2p}{\partial z^2}=0,
\label{eq:axisymdarcy}
\end{equation}
where $p$ is the pressure, $\mathbf u$ is the
superficial Darcy velocity, and $r$ and $z$ denote radial and axial position in $0<r<R$ and
$0<z<\delta$. Pressure is prescribed
on each port disk, the remaining faces have zero normal flux, and the
solution is regular at the axis. Integrating the Darcy velocity over a
port gives $Q$ for the prescribed pressure difference; normalization by
the uniform-feeding resistance in Eq.~\eqref{eq:capacity} then gives $F$.
The constant factor $k/\mu$ cancels from this resistance ratio, so $F$
depends on $a_1/R$, $a_2/R$ and $\delta/R$ under these boundary conditions.

Equation~\eqref{eq:capacity} combines the spreading factor $F$ and
the permeability $k$.
The axisymmetric calculation determines the spreading factor, and we next estimate
$k$~from the pore resistance (Sec.~\ref{sec:perm}).

\subsection{Permeability from pore resistance}\label{sec:perm}
The permeability $k$ is estimated from the resistance of the resolved pore channels.
Porosity is defined as $\phi=V_{\mathrm{void}}/V_{\mathrm{total}}$,
where $V_{\mathrm{void}}$ is the fluid-filled void volume and
$V_{\mathrm{total}}$ is the total plug volume.
Treating the plug as a
bundle of channels of porosity $\phi$, effective width $w_{\rm eff}$ and
tortuosity $\mathcal T=S/L_c$ (arc length over end-to-end length), and
correcting the lubrication resistance by the ratio $C_{\rm CFD}$ of
simulated to lubrication resistance, both based on the section-averaged
pressure drop over the channel, gives
\begin{equation}
k=\frac{\phi\,w_{\rm eff}^2}{12\,\mathcal T^2\,C_{\rm CFD}},\qquad
w_{\rm eff}=\left(\frac{S}{\int_0^S w^{-3}\,\mathrm ds}\right)^{1/3}.
\label{eq:permeability}
\end{equation}
Here $w$ is the local channel width\ at arc length $s$. The $w^{-3}$ weight makes the narrowest throats dominate. The closure is
analogous to classical capillary-bundle models~\cite{carman1997} and
yields an effective, channel-informed permeability rather than a measured
property of a particular plug. The pore simulations in the traced channels give the
pressure drop, and hence $C_{\rm CFD}$ and $k$. Because $C_{\rm CFD}$ comes
from simulation, we next ask
which geometric features set the pore resistance, and whether it can be
predicted without simulation. The theory below is tested on a different
observable, the pressure difference along an interior reference curve, so
its agreement with the simulations explains the resistance but does not
by itself validate a theory-based permeability.

\begin{figure}[!htb]
\centering\includegraphics[width=\textwidth]{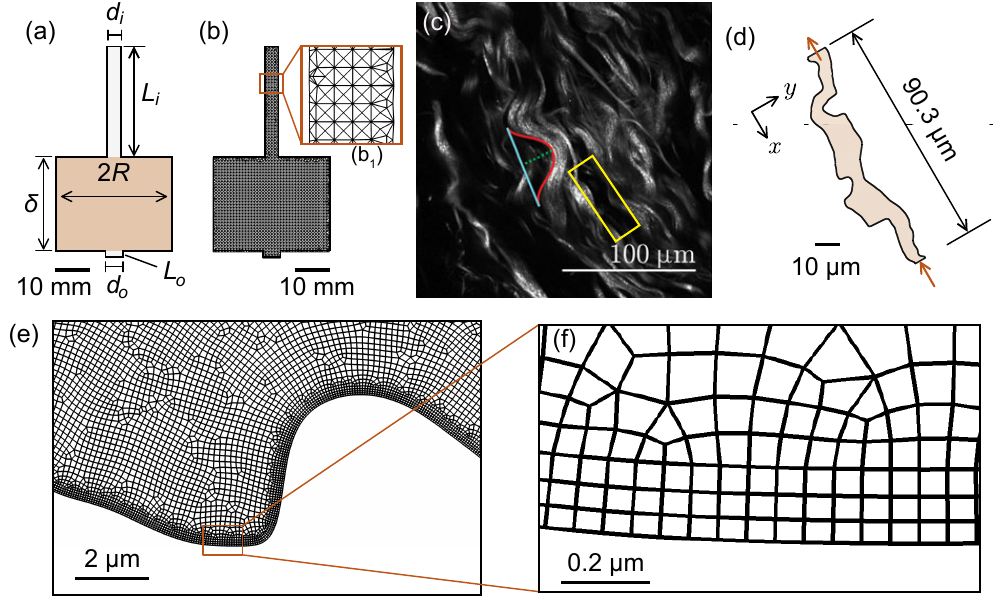}
\caption{\label{fig:geometry}Geometry and mesh from the bioreactor to a
single pore channel. (a) Device cross-section; the shaded chamber holds
the collagen plug. (b) Device mesh, with the boxed region of the inlet
conduit enlarged in (b$_1$). (c) Second-harmonic-generation image
of adventitial collagen, reproduced with permission from J.~A. Niestrawska, A.~Pukaluk,
A.~R. Babu, and G.~A. Holzapfel, ``Differences in collagen fiber diameter
and waviness between healthy and aneurysmal abdominal aortas,''
Microsc.\ Microanal.\ \textbf{28}, 1649--1663 (2022), Fig.~4, under
CC~BY~4.0~\cite{niestrawska2022}; annotations are part of the source
image. (d) Channel G3, the reference of the five channels G1--G5, traced from the dark interstitial space in (c);
arrows give the channel axes and the flow direction, and the dimension
line gives the end-to-end length $L_c$ (90.3~$\mu$m). (e) Pore mesh, with the boxed region
enlarged in (f) to show the wall layers.}
\end{figure}

\subsection{Resistance of a curved pore}\label{sec:curved}
Channel width, curvature and wall offset affect the pressure drop
predicted along the reference curve. Arc length
$s$\ along the smooth reference curve and normal distance $n$
from it define the coordinates, with the signed curvature
$\kappa$\ entering the metric factor $\mathcal H=1-\kappa n$. The walls lie at $n=c(s)\pm h(s)$,
where $h$ is the half-width and $c$ the offset of the wall pair from the
reference curve. The coordinates are locally regular when
$1-\kappa c-|\kappa|h>0$ across the channel. We take $s$ to increase from
the outlet to the inlet, as the physical channel coordinate $X$ does, and
let $q>0$ be the throughflow per unit depth, which moves toward decreasing
$s$. The tangential and normal velocity components
$u_s$ and $u_n$ are related to the streamfunction $\psi$ by
$u_s=-\partial_n\psi$ and $u_n=\mathcal H^{-1}\partial_s\psi$.
Steady Stokes flow~\cite{batchelor1967}
requires $\mathcal L(\mathcal L\psi)=0$, where $\mathcal L$ is the scalar
Laplacian in these coordinates, with $\psi=\pm q/2$ and
$\partial_n\psi=0$ on the walls. For slow variation along the channel we
expand $\psi$ to second order in the ratio of width to the length scale
of wall variation (Sec.~\ref{S-sec:derivation} of the Supplemental
Material~\cite{supplemental}). The pressure gradient on the reference curve $n=0$ and the
pressure drop along it over an interval $s_1<s<s_2$ are then
\begin{align}
\frac{\mathrm dp}{\mathrm ds}&\simeq\frac{3\mu q}{2h^3}(1+\mathcal C),\qquad
\Delta p_{\rm th}=\frac{3\mu q}{2}\int_{s_1}^{s_2}\frac{1+\mathcal C}{h^3}\,\mathrm ds,
\nonumber\\
\mathcal C&=-c\kappa+c_s^2+\tfrac75h_s^2-\tfrac1{10}hh_{ss}
+\tfrac2{15}h^2\kappa^2-cc_{ss}\nonumber\\
&\quad+\frac{6cc_sh_s}{h}-\frac{6c^2h_s^2}{h^2}+\frac{3c^2h_{ss}}{2h},
\label{eq:theory}
\end{align}
where $\mathcal C$ is the geometric correction
to the leading-order gradient and subscripts denote $s$ derivatives.
With $\mathcal C=0$,
Eq.~\eqref{eq:theory} is the leading-order lubrication law
$\mathrm dp/\mathrm ds=12\mu q/w^3$ with $w=2h$. The term $-c\kappa$ is of
first order and the rest of second order. Two limits check the result. For
a flat wall it reduces to the extended-lubrication gradient of Tavakol
\textit{et al.}~\cite{tavakol2017}; for a straight channel tilted by
$\theta$ relative to the reference line, it
reproduces, to second order, the exact relative gradient
$\sec^2\theta$ on $n=0$, whereas the gradient along the wall-pair midline
is $\sec^4\theta$. Because the pressure thus varies across the channel at
second order, the predicted drop is compared with the pore simulations on
the same line and interval, using $h$, $c$ and $\kappa$ taken from the same
traced channels. Curvature and offset carry the geometric content that a
fixed-axis expansion would treat as large wall slopes.

\subsection{Internal and outlet replacement}\label{sec:replacement}
Pressure fixes how much flow the plug can take; it does not say how
quickly that flow replaces the fluid inside. In a resolved channel of
area $A_{2D}$, one ideal flush takes $\tid=A_{2D}/q$. If $t_{99}$ is the
time to reach 99\% fresh fluid, the replacement efficiency relative to an
ideal flush is
\begin{equation}
\mathcal E_{99}=\frac{\tid}{t_{99}},\qquad \tid=\frac{A_{2D}}{q}.
\label{eq:efficiency}
\end{equation}
The same comparison underlies the original design estimate of the device,
in which an inlet volume $Q\Delta t$\,delivered
over time $\Delta t$, with $Q=\pi d_i^2\bar u_i/4$, is compared with the
plug pore volume $\phi\pi R^2\delta$; once $Q$ is fixed
the inlet diameter does not enter separately. Here
$d_i$ is the inlet diameter and $\bar u_i$ is the mean inlet speed. A passive solute of
concentration $c$ and diffusivity $D$, carried by the steady velocity
$\mathbf u$, obeys
\begin{equation}
\partial_t c+\mathbf u\cdot\nabla c=D\nabla^2c,
\label{eq:tracer}
\end{equation}
with $c=0$ initially and $c=1$ at the inlet. We record two replacement
times: $t_{99,o}$, when the flux-weighted outlet concentration
$\bar c_o$ reaches 0.99, and $t_{99,V}$, when the
area-averaged internal concentration $\bar c_V$ does.

The two observables differ fundamentally. Without diffusion, a streamline
carrying flux $\mathrm dq$ with transit time $T$ holds area $T\,\mathrm dq$,
so the outlet and internal concentrations after a unit step are
\begin{equation}
\bar c_o(t)=\frac1q\int_{T\leq t}\mathrm dq,\qquad
\bar c_V(t)=\frac1{A_{2D}}\int\min(t,T)\,\mathrm dq.
\label{eq:advectivelimit}
\end{equation}
Equation~\eqref{eq:advectivelimit} implies no universal ordering: in
uniform plug flow the interior reaches 99\% at $0.99\tid$, before the
outlet. In a pore, however, slow near-wall streamlines hold area but carry
little flux, so they can delay internal replacement, and closed
recirculation is never reached without diffusion. Diffusion across
streamlines shortens such delays at a rate set by
$\Lambda_\perp=D\tid/w_{\rm eff}^2$. Where a pocket exchanges with the
main passage only slowly, the internal residual $r=1-\bar c_V$ decays at
late times as $r\simeq B_0\,e^{-(t-t_0)/\te}$,
where $B_0$ is the fitted amplitude and $\te$
is the exchange time, with $t$ measured from the
inlet switch and $t_0$ the start of the fitted tail, consistent with mobile--immobile
exchange~\cite{coats1964,danckwerts1953}, and the exchange number
$\Lambda_e=\te/\tid$ compares exchange with one flush. We test whether
$\Lambda_e$ orders the ratio $t_{99,V}/t_{99,o}$ rather than assuming a
threshold.
The two residuals are also linked by conservation of solute. With
the inlet held at $c=1$ and no diffusive flux at the outlet, the internal
average obeys
\begin{equation}
\tid\,\frac{\mathrm d\bar c_V}{\mathrm dt}=1-\bar c_o,
\label{eq:balance}
\end{equation}
apart from diffusion through the inlet, which is negligible once the
inlet region is fresh; this is the washout relation of residence-time
theory~\cite{danckwerts1953}. In a single-mode tail,
Eq.~\eqref{eq:balance} gives $1-\bar c_o=r/\Lambda_e$: the interior
residual is $\Lambda_e$ times the outlet residual. A threshold crossed
within that tail is therefore reached later in the interior than at the
outlet when $\Lambda_e>1$, and earlier when $\Lambda_e<1$. The tracer simulations give $t_{99,o}$, $t_{99,V}$ and $\te$,
which are interpreted with Eq.~\eqref{eq:advectivelimit} and $\Lambda_e$.
Flow strength is measured by the P\'eclet number
$\Pe=(q/\bar w)w_{\rm eff}/D$, with $\bar w$ the mean width. The
dimensionless groups are collected in Table~\ref{tab:groups}.

\begin{table}[H]
\caption{\label{tab:groups}Dimensionless groups used in this study. 
$R_{\rm plug}=\Delta p_{\rm plug}/Q$ is the hydraulic resistance of the plug
and $\mathrm{Sc}=\mu/(\rho D)$ the Schmidt number, where $\rho$ is the fluid density.}
\setlength{\tabcolsep}{3pt}
\begin{ruledtabular}
\begin{tabular}{lll}
Group & Definition & Meaning\\
$F$ & $R_{\rm plug}kA/(\mu\delta)$ & Resistance increase from port feeding\\
$\Pi_P$ & $\Delta p_{\rm plug}/\Delta P_c$ & Fraction of supply pressure used\\
$\mathcal E_{99}$ & $\tid/t_{99}$ & Replacement efficiency\\
$\Pe$ & $(q/\bar w)w_{\rm eff}/D$ & Advection versus diffusion\\
$\mathrm{Re}_w$ & $\Pe/\mathrm{Sc}$ & Inertia versus viscosity in a pore\\
$\Lambda_\perp$ & $D\tid/w_{\rm eff}^2$ & Transverse diffusion per flush\\
$\Lambda_e$ & $\te/\tid$ & Pocket exchange time per flush\\
\end{tabular}
\end{ruledtabular}
\end{table}

\section{Models and methods}
Five calculations address the \hyperlink{intro-q1}{pressure} and \hyperlink{intro-q2}{replacement} questions (Sec.~\ref{sec:introduction})
at the device, plug and pore scales. Two simulations use the three-dimensional geometry
of one chamber and its inlet and outlet conduits. In the \emph{open-device simulations}, the chamber holds
fluid only, and steady laminar flow gives the pressure difference across
the housing and the speed at its outlet. In the \emph{three-dimensional
porous-plug simulation}, the chamber is filled with a porous zone
representing the plug, and steady Darcy--Stokes flow gives the pressure
difference across the plug and hence $F_{3D}$. At the pore scale, two calculations use the same two-dimensional geometry. The \emph{pore simulations} solve steady creeping
flow between rigid collagen walls in the channels G1--G5~(Table~\ref{tab:geometry}) and give the pore
resistance $C_{\rm CFD}$ and the permeability $k$. The \emph{tracer
simulations} follow a passive scalar in time on the frozen pore velocity
in G1, G3 and G5 and give $t_{99,o}$, $t_{99,V}$ and $\te$. The fifth, the
\emph{axisymmetric Darcy calculation}, solves Laplace's equation in a
homogeneous plug and gives the spreading factor $F$ of
Eq.~\eqref{eq:capacity}. The device geometry is that of the bioreactor
described in Sec.~\ref{sec:experiment}.
The modeled flow through the plug ranges from 5 to 30~mL/min.
\subsection{Open-device simulations}
The open-device calculation isolates the housing loss by solving steady laminar flow in one fluid-filled chamber and its inlet and outlet conduits. The domain (Fig.~\ref{fig:geometry}(a,b)) has an inlet conduit of
diameter $d_i=3.86$~mm and length $L_i=31.7$~mm, a chamber of radius
$R=16.66$~mm and height $\delta=26.87$~mm, and an outlet of diameter
$d_o=5.00$~mm and length $L_o=1.78$~mm, all as meshed.
The fluid has constant density $\rho=1012$~kg/m$^3$ and viscosity
$\mu=9.64\times10^{-4}$~Pa\,s, prescribed rather than measured. Steady
laminar flow obeys
\begin{align}
&\nabla\cdot\mathbf u=0,\qquad
\rho\,\mathbf u\cdot\nabla\mathbf u=-\nabla p_m+\mu\nabla^2\mathbf u,\nonumber\\
&p_m=p-\rho\,\mathbf g\cdot\mathbf x,
\label{eq:ns}
\end{align}
where $p$ is the physical pressure and $p_m$ the modified pressure, $\mathbf g$ is gravitational acceleration, and $\mathbf x$ is position. Gravity
is enabled with the solver's automatic operating density. For this constant-density fluid the setting removes the hydrostatic contribution, leaving $p_m$ as the solved pressure. Walls are no-slip, and pressure inlet and outlet values
deliver the target flow. We used the finite-volume solver ANSYS Fluent
2025 R2~\cite{ansys2025r2} for the open-device, pore, tracer and porous-plug simulations.

The flow and tracer calculations use the same finite-volume framework.
Equations~\eqref{eq:ns} and \eqref{eq:tracer} are integrated over each
control volume $V_P$ with faces $f$ of area $A_f$ and outward normal
$\mathbf n_f$, giving the discrete balances
\begin{align}
&\sum_f\rho(\mathbf u_f\!\cdot\!\mathbf n_f)\mathbf u_fA_f=
-\sum_fp_{m,f}\mathbf n_fA_f+\sum_f\mu(\nabla\mathbf u)_f\!\cdot\!\mathbf n_fA_f,\nonumber\\
&V_P\,\delta_t c_P+\sum_f(\mathbf u_f\!\cdot\!\mathbf n_f)c_fA_f
=\sum_fD(\nabla c)_f\!\cdot\!\mathbf n_fA_f,
\label{eq:fv}
\end{align}
with $\sum_f\mathbf u_f\!\cdot\!\mathbf n_fA_f=0$ for mass, where $\delta_t$
is the discrete implicit backward-difference time derivative.
Equation~\eqref{eq:fv} is
the constant-density form for the open fluid, and the porous zone of the
three-dimensional tracer simulation adds the porosity to the storage term, $\phi V_P\,\delta_t c_P$. The tracer simulations
use the bounded second-order implicit formulation
(Sec.~\ref{S-sec:numerics}). Convective face
values are interpolated with a second-order upwind scheme, face pressure
to second order, and diffusive fluxes use reconstructed face gradients
with non-orthogonal correction. Pressure and velocity are coupled with the
SIMPLEC algorithm in the open-device simulations~\cite{patankar1980,versteeg2007}.
Iterations stop when the continuity and velocity residuals fall below
$10^{-4}$ and $10^{-5}$. The housing loss $\Delta p_{\rm dev}$ is the area-weighted total-pressure difference between the open-device inlet and outlet (Sec.~\ref{S-sec:numerics}). Three grids of about 0.5, 1 and 2 million cells
bound its grid sensitivity.

\subsection{Pore geometry}
The pore calculations use channels traced from the dark interstitial space of a
second-harmonic-generation image of healthy aortic adventitial
collagen~\cite{niestrawska2022} (Fig.~\ref{fig:geometry}(c,d)). These channels represent dense collagen pore space and are not reconstructions of the reactor plug.
Five intensity thresholds (the 28th to 45th percentiles) give channels
G1--G5 (Table~\ref{tab:geometry}). With G3 as the reference, the channels from this single image test sensitivity to segmentation rather than biological variation.

\begin{table}[H]
\caption{\label{tab:geometry}Pore channels traced from the collagen image
and their permeabilities at assumed porosity $\phi=0.27$. Widths from a
2~$\mu$m-smoothed reference curve; $w_{\rm in}$ is the meshed
pore opening at the inlet $X=L_c$. Lengths in $\mu$m, area in
$\mu$m$^2$, permeability in $10^{-13}$~m$^2$.}
\begin{ruledtabular}
\begin{tabular}{lrrrrrrrrr}
Channel & Threshold percentile & $L_c$ & $A_{2D}$ & $\mathcal T$ & $w_{\mathrm{eff}}$ &
$\bar w$ & $w_{\rm in}$ & $C_{\mathrm{CFD}}$ & $k$\\
G1 & 28 & 71.33 & 478.0 & 1.052 & 4.53 & 6.24 & 2.715 & 1.323 & 3.15\\
G2 & 31 & 73.22 & 537.9 & 1.063 & 4.90 & 6.81 & 2.708 & 1.285 & 3.72\\
G3 & 35 & 90.30 & 768.4 & 1.091 & 6.11 & 7.66 & 4.609 & 1.269 & 5.57\\
G4 & 40 & 94.47 & 880.3 & 1.081 & 7.00 & 8.48 & 4.825 & 1.252 & 7.55\\
G5 & 45 & 96.75 & 972.7 & 1.078 & 7.81 & 9.15 & 12.19 & 1.277 & 9.24\\
\end{tabular}
\end{ruledtabular}
\end{table}

Both the permeability closure and the theory comparison use walls smoothed with a Gaussian filter of standard deviation
0.75~$\mu$m. The 2~$\mu$m reference curve of Table~\ref{tab:geometry}
supplies the hydraulic width and tortuosity of the permeability closure.

The theory comparison uses a more strongly smoothed reference curve
because the local metric fails on part of each channel at 2~$\mu$m (Table~\ref{S-tab:metric}).
We chose 4~$\mu$m for the reference after checking that the local metric stays positive on every comparison interval. The reference smoothing was not selected prospectively. The 3~$\mu$m curve also meets this condition and serves as a sensitivity check. For the correction $\mathcal C$, $h$, $c$
and $\kappa$ were filtered at 1~$\mu$m before differentiation, whereas the
leading $h^{-3}$ term uses the unfiltered constructed half-width. Because
this regularization is a choice rather than part of the theory, we also
vary the filter width and apply the filtered half-width to both terms
(Table~\ref{S-tab:theoryranges}). Four
pairs of end cuts define the comparison intervals, 4 and 8, 6 and 12,
8 and 16, and 10 and 20~$\mu$m at the two ends; the reference case
excludes 6 and 12~$\mu$m (Table~\ref{S-tab:theoryranges}), which
also gives the sensitivity of the theory to the reference smoothing and end
cuts.

The 1~$\mu$m filter enters only the theory. The permeability depends
on the reference curve only through the slot area $Sw_{\rm eff}$.
Since the leading-order lubrication pressure drop is $12\mu qS/w_{\rm eff}^3$,
Eq.~\eqref{eq:permeability} reduces to
$k=\phi\mu qL_c^2/(Sw_{\rm eff}\,\Delta p_{\rm CFD})$, with $\Delta p_{\rm CFD}$
the simulated pressure drop at flow $q$ per unit depth. Replacing this area
by the traced area $A_{2D}$, which does not depend on the choice of
reference curve, tests the area convention (below).

The reference porosity is adopted from the collagen-scaffold study
of Sawadkar \textit{et al.}~\cite{sawadkar2021}. They quantified porosity
from segmented and binarized three-dimensional micro-CT reconstructions
using CTAn software and reported 27.33\% for collagen. Converting this
percentage to a fraction gives $\phi=27.33/100=0.2733$. We use the rounded
value $\phi=0.27$ as the model input for all five channels.
The scaffold study supplies the porosity, and the collagen image supplies the pore geometry. In that image the dark-area fraction at the five thresholds is
about 0.36--0.53, which for an isotropic section estimates the
corresponding volume fraction~\cite{underwood1970}. It is an upper
estimate of the pore space, because dark regions also contain
non-collagenous tissue. Used in place of 0.27, these fractions would raise
$k$ by factors of 1.3--1.9 and lower the port-fed demand at 5~mL/min from
about 32 to about 20~kPa for G3. The pressure-budget conclusions are
unchanged. Equation~\eqref{eq:permeability} also assigns each channel the
pore volume of a uniform slot of width $w_{\rm eff}$. Assigning it the
traced area $A_{2D}$ instead multiplies $k$ by $Sw_{\rm eff}/A_{2D}$, which
is 0.71--0.84 for G1--G5, and raises the G3 demand to about 40~kPa, a
shift comparable in magnitude to the effect of porosity, in the opposite direction.
Porosity can exceed 0.99~\cite{avendano2020} in dilute reconstituted gels.
Since $k\propto\phi$ at fixed geometry, the
absolute pressure demand scales as $1/\phi$ and requires a measured
permeability before it is applied to a specific plug.

\subsection{Pore and tracer simulations}
The pore flow fields give the hydraulic resistance and supply the velocities for tracer transport. The simulations solve Eq.~\eqref{eq:ns} without gravity in the
two-dimensional channels G1--G5, with no slip on the
walls, a uniform inlet velocity at $X=L_c$ and zero pressure at the outlet
$X=0$.
The pore and open-device simulations share the same spatial discretization.
The pore simulations use SIMPLE coupling and converge to residuals of about $10^{-8}$. The channel Reynolds number stays below
0.51, and pressure scales linearly with flow within 0.02\%, so the
steady velocity field was frozen and reused for transport. In the tracer
simulations, Eq.~\eqref{eq:tracer} was solved on these frozen fields as a
user-defined scalar with
second-order upwind convection, $D=6.7\times10^{-10}$~m$^2$/s (entered as
$\Gamma=\rho D=6.78\times10^{-7}$~kg\,m$^{-1}$\,s$^{-1}$) and a time step of
$\tid/2000$ for the runs in Table~\ref{tab:tracer}, with bounded
second-order implicit time integration.
Twelve tracer simulations (Table~\ref{tab:tracer}) span
$\Pe=3.6$--729: seven in G3 (E1--E7), three in G1 (E8--E10) and two in G5
(E11, E12).

Two mappings translate device flow into a pore inlet speed for the tracer runs. The conduit mapping sets it equal to the peak speed
at the open-device outlet. The Darcy mapping matches the pore flush time to transit over the end-to-end length $L_c$ at the plug-scale interstitial
speed $Q/(A\phi)$, giving
\begin{equation}
u_{\rm in,D}=\frac{A_{2D}\,Q}{w_{\rm in}L_c\,A\phi},
\label{eq:mapping}
\end{equation}
where $w_{\rm in}$ is the width of the meshed pore inlet at $X=L_c$ (Table~\ref{tab:geometry}). The prescribed speeds are listed in
Table~\ref{tab:tracer}.
At 5~mL/min, the conduit mapping gives pore inlet speeds about 9, 12 and 26
times the Darcy-mapped speeds in G1, G3 and G5, respectively. We therefore report the tracer results against the pore P\'eclet
number $\Pe$ rather than the nominal device flow rate.
Section~\ref{sec:res_plug} relates both mappings to the speeds inside the plug.

\begin{table}[H]
\caption{\label{tab:tracer}Tracer simulations and replacement times. Ch.\ is the pore channel.
E1 and E2 use the Darcy velocity mapping, all others the conduit
mapping. $u_{\rm in}$ is the prescribed uniform inlet speed in mm/s. Times $\tid$ and $\te$ in ms; all values to three significant
figures.}
\setlength{\tabcolsep}{2.4pt}
\begin{ruledtabular}
\begin{tabular}{llrrrrrr}
Run & Ch. & $u_{\rm in}$ & Pe & $\tid$ & $\te$ & $t_{99,o}/\tid$ & $t_{99,V}/\tid$\\
E1 & G3 & 0.653 & 3.58 & 255 & 45.4 & 1.75 & 1.44\\
E2 & G3 & 3.92 & 21.5 & 42.5 & 11.3 & 1.94 & 1.59\\
E3 & G3 & 7.69 & 42.2 & 21.7 & 7.76 & 2.19 & 1.82\\
E4 & G3 & 14.0 & 76.8 & 11.9 & 5.62 & 2.49 & 2.14\\
E5 & G3 & 20.9 & 115 & 7.98 & 4.59 & 2.73 & 2.41\\
E6 & G3 & 29.2 & 160 & 5.71 & 3.87 & 2.94 & 2.68\\
E7 & G3 & 46.9 & 257 & 3.56 & 3.06 & 3.26 & 3.12\\
E8 & G1 & 7.69 & 22.6 & 22.9 & 28.3 & 2.44 & 2.62\\
E9 & G1 & 23.5 & 69.0 & 7.51 & 24.1 & 2.60 & 5.16\\
E10 & G1 & 46.9 & 138 & 3.75 & 21.9 & 2.88 & 8.47\\
E11 & G5 & 7.69 & 119 & 10.4 & 5.01 & 2.60 & 2.24\\
E12 & G5 & 46.9 & 729 & 1.70 & 1.89 & 3.81 & 3.91\\
\end{tabular}
\end{ruledtabular}
\end{table}


\subsection{Axisymmetric Darcy calculation and porous-plug simulation}
The pore calculations provide the permeability.
The Darcy calculation gives the spreading factor $F$ for port feeding from the axisymmetric Laplace problem of Sec.~\ref{sec:darcy}
with the device port radii $a_1=1.93$~mm and $a_2=2.50$~mm, using an
in-house finite-volume solver (equipotential port disks,
all other plug faces sealed) on three grids. The code is provided with the
data.
With this reduced model we can also vary port size, as reported in Sec.~\ref{sec:res_pressure}, to isolate the effect of feeding geometry from material resistance. The discretization and normalization are given in Sec.~\ref{S-sec:darcy_solver} of the Supplemental Material.
Inside the plug, the Darcy face fluxes give the Stokes stream function $\psi$, which
rises from 0 on the axis to $Q$ on the side wall, and the superficial
speed relative to the uniform value $Q/A$. The advective transit time of
the stream tube between $\psi$ and $\psi+\mathrm d\psi$ follows from the
pore volume it holds,
\begin{equation}
T(\psi)=\phi\,\frac{\mathrm dV}{\mathrm d\psi},
\label{eq:transit}
\end{equation}
where $V(\psi)$ is the plug volume enclosed by the stream surface $\psi$.
The transit times give outlet and interior replacement at the plug scale through
Eq.~\eqref{eq:advectivelimit}, with $\mathrm dq$ replaced by
$\mathrm d\psi$ and $A_{2D}$ by the pore volume, and time measured in mean
residence times $\bar t=\phi\pi R^2\delta/Q$, so the scaled results do not
depend on $k$, $\mu$, $Q$ or $\phi$. The calculation is advective. Dispersion within the plug is not modeled, and Sec.~\ref{sec:res_plug}
compares it with a three-dimensional simulation that includes diffusion.
The pressure calculation is checked by resolving the porous plug and conduits together in the three-dimensional model of one chamber. The chamber is filled with an isotropic porous zone
of permeability $k=5.57\times10^{-13}$~m$^2$ (G3), viscous resistance
$1/k=1.80\times10^{12}$~m$^{-2}$, assumed porosity 0.27 and zero
inertial resistance, using the superficial-velocity formulation, so that
the momentum equation gains the sink $-(\mu/k)\mathbf u$. The remaining plug
faces are sealed, the open inlet and outlet conduits are retained, the
outlet is held at a reference pressure, and gravity is omitted. Two runs
at the nominal 5~mL/min prescribe a uniform inlet velocity of
$7.12\times10^{-3}$~m/s, which delivers 4.99~mL/min on the meshed inlet
disk. A third prescribes a mass flow of $5.06\times10^{-4}$~kg/s
(30.0~mL/min) to test linearity. Table~\ref{tab:porous3d} gives the spreading factor $F_{3D}=\Delta p_{\rm disks}kA/(\mu\delta Q)$ with the delivered $Q$ and the area-weighted pressure difference $\Delta p_{\rm disks}$
between the port disks. We also compute the flow-weighted difference for comparison.
The production mesh has 630\,657 tetrahedral cells and a port-edge size of
0.10~mm, and a port-refined mesh with 1\,385\,111 cells and a port-edge
size of 0.05~mm tests mesh sensitivity. Their minimum orthogonal quality
is 0.15 and 0.13. The velocity residuals fell below $10^{-8}$ while the
scaled continuity residual leveled off near $5\times10^{-6}$, and the
mass imbalance was below $3\times10^{-7}$ of the inlet flow in every run.
For comparison with the advective limit, the plug-scale calculation solves the tracer equation, Eq.~\eqref{eq:tracer}, on the frozen superficial-velocity flow of the production run,
with the same diffusivity in all zones (no porosity factor on the diffusive flux), the porosity of 0.27 in its storage term, a
unit step at the inlet, the schemes of the pore simulations and a time
step of 0.05~s for 420~s, about 5.5 mean residence times. The outlet
concentration and the plug average are referred to the plug by
subtracting the mean transit times of the conduits.
The plug average is shifted by the inlet conduit transit time, 4.45~s,
and the device outlet by the transit times of both conduits, 4.45 and 0.42~s. The shift removes their means
but not their laminar spreading. Details are given in Sec.~\ref{S-sec:r1t}.
\begin{figure}[!htb]
\centering\includegraphics[width=\textwidth]{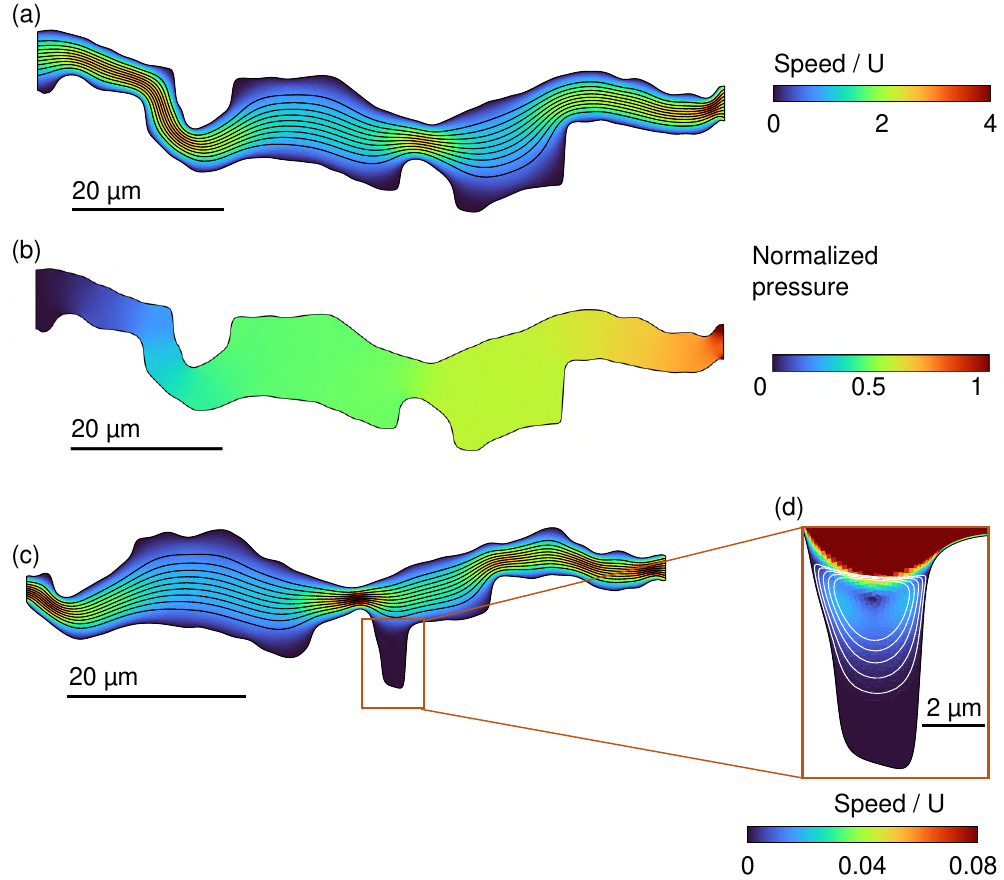}
\caption{\label{fig:pore}Narrow throats control flow in the pore.
G1 and G3 are two of the five pore channels G1--G5 traced from the
collagen image. (a) In channel G3, the reference channel, speed peaks at the throats and streamlines follow the
channel. (b) Pressure falls mainly across the throats and is nearly
uniform in the wide sections; values above 1 at the inlet corners,
where uniform inflow meets the no-slip wall, are shown at the top of the
scale. (c) Channel G1 contains a side pocket (box). (d) Flow in the
pocket, shown on a 50 times finer speed scale, recirculates in closed
loops (white contours); speeds above $0.08U$, including the main passage,
are shown at the top of the scale. Speed is scaled by the mean speed
$U=qL_c/A_{2D}$ and pressure by the inlet-to-outlet difference.}
\end{figure}
\subsection{Verification}
Pressure sensitivity to grid spacing is assessed with the grid convergence index~\cite{celik2008,roache1997} on three G3 grids (21\,295, 37\,045 and
52\,059 cells, grids B3, B2 and B1; the pore mesh is shown in Fig.~\ref{fig:geometry}(e)). With representative spacing $h_g\propto N^{-1/2}$, where $N$ is the cell count, the refinement ratio is
$r_{ij}=h_{g,i}/h_{g,j}=(N_j/N_i)^{1/2}$.
The refinement ratios are unequal, $r_{21}=1.185$ and $r_{32}=1.319$; the
first is below the preferred minimum of about 1.3~\cite{celik2008}, so the
index, which is reported for the pressure, is indicative. For a
quantity $f$ on the fine, medium and coarse grids
($f_1,f_2,f_3$, with $\varepsilon_{ij}=f_i-f_j$),
the observed order $p_o$ and the fine-grid index follow from
\begin{equation}
\frac{\varepsilon_{32}}{\varepsilon_{21}}=\frac{r_{21}^{p_o}(r_{32}^{p_o}-1)}{r_{21}^{p_o}-1},\qquad
\mathrm{GCI}=\frac{1.25\,|1-f_2/f_1|}{r_{21}^{p_o}-1},
\label{eq:gci}
\end{equation}
evaluated with the unrounded ratios.
At all four interior stations the index is below 0.3\%, and halving the
first wall-cell height of the near-wall layers (Fig.~\ref{fig:geometry}(f)) changes the pressure by less than 0.08\%.
Station-by-station values are given in the Supplemental Material
(Fig.~\ref{S-fig:s_pressure}, Table~\ref{S-tab:verification}).

For the tracer, an independent finite-volume scalar solver uses the same frozen fluxes and reproduces the Fluent replacement times to about 0.5\%,
although it uses central advection with second-order backward
differencing in time, whereas Fluent uses second-order upwind convection.
With that solver, the outlet and interior 99\% times of E3 and E7
change by less than 0.6\% over four G3 grids of 21\,295--59\,431 cells,
the finest with refined near-wall layers.

The pocket channel G1 needs a separate check, because its lag is set
by slow exchange across the pocket mouth, where numerical diffusion could
act. On the original G1 mesh the two solvers give lag ratios
$t_{99,V}/t_{99,o}$ of 2.95 and 2.94 in E10, and its 99\% times are unchanged to six digits between 1000 and
4000 time steps per flush time (Fig.~\ref{S-fig:s_tracer}(a)). As a two-mesh
sensitivity check, we then repeated E8 and E10 on a G1 mesh with four
times as many cells, made by splitting every cell into four. The 99\%
times change by at most 2.4\%, the lag ratio of E10 moves from 2.94 to
2.99 and the exchange time changes by less than 2\%
(Table~\ref{S-tab:g1refine}). The lag persisted after refinement and was
slightly larger on the finer mesh. Because the refined runs began with
smaller time steps, these differences describe the two configurations
tested rather than a formal estimate of the spatial error. The small
lag in G5 run E12 remains a result on its single mesh.

\section{Results}
\subsection{What sets the pore resistance}

To answer Q1, we start with the constrictions that control the pore resistance. The plug permeability in Eq.~\eqref{eq:permeability} depends on that resistance. In channel
G3 the flow accelerates through each narrow throat, where streamlines
crowd together, and the pressure falls mainly across those throats while
remaining nearly uniform in the wide sections (Fig.~\ref{fig:pore}(a,b)).
Channel G1 contains a side pocket in which the flow is slow and
recirculates in closed loops (Fig.~\ref{fig:pore}(c,d)). The effect of this pocket on replacement is examined in Sec.~\ref{sec:res_repl}.

\begin{figure}[H]
\centering\includegraphics[width=\textwidth]{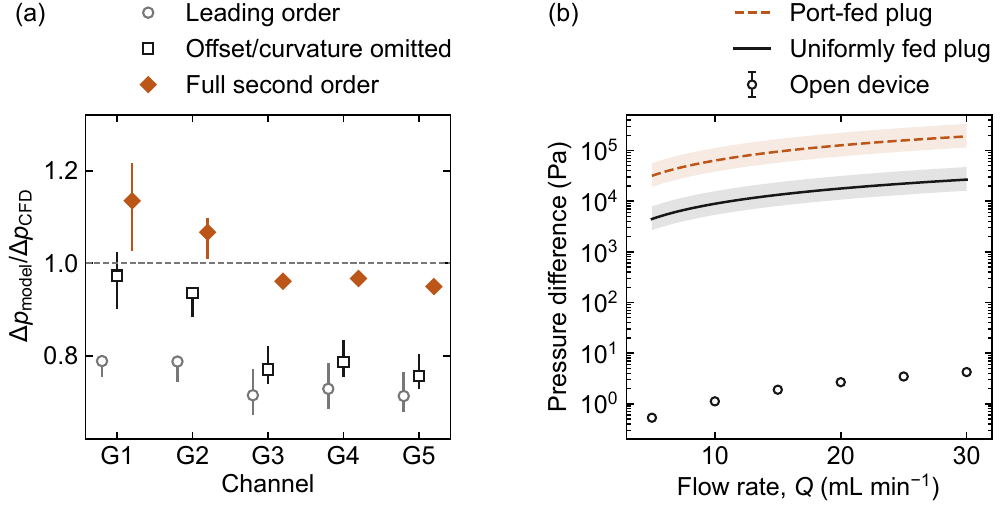}
\caption{\label{fig:pressure}Pore-resistance predictions and the
pressure budget of the bioreactor. (a) Ratio of predicted to simulated
pressure drop for the traced channels G1--G5 (Table~\ref{tab:geometry}):
leading-order lubrication (circles), second-order theory without
curvature and offset terms (squares), and the full second-order theory
(diamonds). Markers show the reference case (4~$\mu$m reference smoothing,
6 and 12~$\mu$m excluded at the ends); bars span both regular smoothings
(3 and 4~$\mu$m) and all four comparison intervals and are parameter
ranges, not confidence intervals. The dashed line marks exact agreement.
(b) Pressure difference across the open device (circles; bars span three
grids that are not a systematic refinement family) and pressure demand
of a sealed homogeneous plug with uniform feeding (solid) and port
feeding (dashed, $F=7.12$), using the G3 permeability at $\phi=0.27$.
Bands span G1--G5 and show segmentation sensitivity from one image,
not statistical uncertainty. For the assumed permeability, the plug
pressure demand exceeds the device pressure loss.}
\end{figure}

Narrowing and turning occur together at the throats, where a straight-channel estimate is expected to miss part of the resistance. Using the section-averaged pressure drop, the simulated resistance exceeds the
leading-order estimate by 25--32\% ($C_{\rm CFD}=1.25$--1.32,
Table~\ref{tab:geometry}); the theory comparison below uses the interior
reference-curve pressure instead.
Figure~\ref{fig:pressure}(a) tests whether the curved-channel theory of
Sec.~\ref{sec:curved} recovers this missing resistance from the channel
shape alone. For each channel, the pressure drop predicted by
Eq.~\eqref{eq:theory} is divided by the simulated drop over the same
stretch of the reference curve. Leading-order lubrication underpredicts by
roughly 20--33\% in all channels. For G3--G5, keeping the second-order
terms but dropping both curvature and offset still underpredicts by
17--27\%, whereas the full second-order theory comes within 3--6\%, for
both regular reference smoothings and all comparison intervals
(Fig.~\ref{fig:pressure}(a)). This close agreement relies on the
1~$\mu$m regularization of the derivatives. For filter widths of
0.5--1.5~$\mu$m, or with the filtered half-width also in the leading term,
the theory stays within about 13\% of the simulations, still far closer
than leading order (Table~\ref{S-tab:theoryranges}). Most of the
second-order gain comes from the offset terms, and curvature adds about
4--5 percentage points. The reference interval holds about 64--80\% of
the leading-order resistance $\int w^{-3}\,\mathrm ds$ in G3--G5 and about
40--50\% in G1 and G2, because narrow sections near the channel ends
carry a large share of it. For the narrower channels G1 and G2, whose walls
are steeper, the full theory overpredicts by up to about 22\%.

Even where the theory agrees with CFD, the correction is appreciable. Over the 2--4~$\mu$m reference smoothings its
resistance-weighted mean in G3--G5 is 20--33\% of the leading term. The agreement is thus empirical. The theory provides a geometry-based estimate of the resistance and explains how throat width and channel turns act together. For the pressure budget we use the permeability inferred from the simulations.

\subsection{Pressure needed to perfuse the plug}\label{sec:res_pressure}

Using the permeability inferred from the pores, we compare the pressure drop across the housing with the pressure required to drive the same flow through the plug.
Figure~\ref{fig:pressure}(b) shows both against flow rate. The open-device pressure difference is only 0.5--4.3~Pa over 5--30~mL/min,
whereas the plug, with the G3 permeability of $5.57\times10^{-13}$~m$^2$
at the assumed porosity, demands several kilopascals even with uniform
feeding, several orders of magnitude above the device pressure
difference.
The flow field of the open device is shown in Fig.~\ref{S-fig:s_device}.

\begin{figure}[h!]
\centering\includegraphics[width=\textwidth]{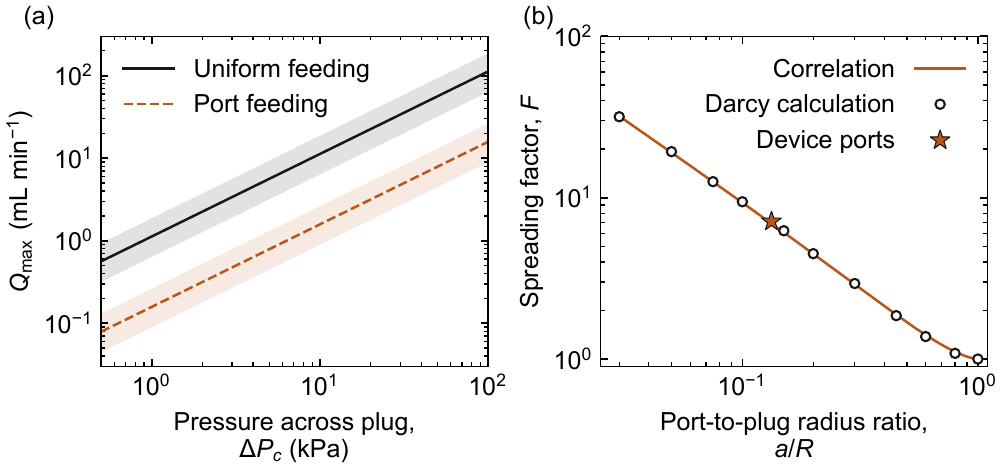}
\caption{\label{fig:capacity}How fluid enters the plug sets its flow
capacity. (a) Largest sustainable flow against the pressure that the
circuit can supply across the plug; feeding through the two ports instead of the whole face
lowers the capacity about sevenfold. Bands span the traced channels G1--G5.
(b) Spreading factor against the ratio of port to plug radius. Smaller
ports force more spreading; the star marks the device ports. The
axisymmetric Darcy calculations (circles) follow a constriction
correlation (line).}
\end{figure}

How the fluid enters and leaves the plug has the largest effect among the modeled factors. Flow spreads from the small inlet disk and converges at the outlet disk, raising the resistance by
$F=7.1$ (Fig.~\ref{fig:capacity}(b)). At 5~mL/min the G3 plug therefore needs
about 4.5~kPa with uniform feeding but about 32~kPa with port feeding, and
the five channels span 19--56~kPa for port feeding (Fig.~\ref{fig:pressure}(b)).
For a given pressure across the plug and a fixed permeability, the
port-fed plug sustains about one-seventh of the flow achieved with uniform
feeding (Fig.~\ref{fig:capacity}(a)).
The numerical $F$ agrees with a classical constriction correlation to
within about 3\%, and its variation over three grids (7.09--7.29) is small
compared with the effect of feeding itself.

The full three-dimensional porous-plug calculation checks the spreading factor for the actual device ports at the prescribed permeability
(Table~\ref{tab:porous3d}). At the nominal 5~mL/min the pressure
difference between the two port disks is about 31.5~kPa, giving
$F_{3D}=7.12$ on the production mesh and 7.10 on the port-refined mesh,
against 7.12 from the axisymmetric calculation. Across the three runs,
pressure drops computed from area- and mass-flow-weighted port averages
differ by less than 0.01\%. At 30~mL/min the pressure difference rises in
proportion to the flow ($F_{3D}=7.12$), as Darcy flow requires. The conduits add only a few
pascals, as in the open-device simulations.

At fixed geometry, viscosity and permeability,
Eq.~\eqref{eq:capacity} also gives approximately 3.2~kPa at 0.5~mL/min
and 6.3~kPa at 1~mL/min through the plug. These are lower-flow predictions
of the same Darcy model, not additional CFD results; a worked evaluation
is given in Sec.~\ref{S-sec:darcy_solver} of the Supplemental Material.

\begin{table}[H]
\caption{\label{tab:porous3d}Three-dimensional porous-plug simulation.
$Q$ is the delivered flow in mL/min, $\Delta p_{\rm disks}$ the
area-weighted pressure difference between the port disks in kPa, and the
imbalance is relative to the inlet mass flow.}
\setlength{\tabcolsep}{3pt}
\begin{ruledtabular}
\begin{tabular}{lrrrrr}
Mesh & $Q$ & Cells & $\Delta p_{\rm disks}$ & $F_{3D}$ & Imbalance\\
Production & 4.99 & 630\,657 & 31.56 & 7.12 & $5\times10^{-8}$\\
Port-refined & 4.99 & 1\,385\,111 & 31.47 & 7.10 & $4\times10^{-8}$\\
Production & 30.0 & 630\,657 & 190.0 & 7.12 & $2\times10^{-7}$\\
\end{tabular}
\end{ruledtabular}
\end{table}

The pressure comparison identifies the dominant resistance within the model assumptions: for the assumed
permeability, the plug, not the device, sets the pressure needed for
perfusion, and the feeding geometry sets most of it.
\begin{figure}[h!]
\centering\includegraphics[width=0.5\textwidth]{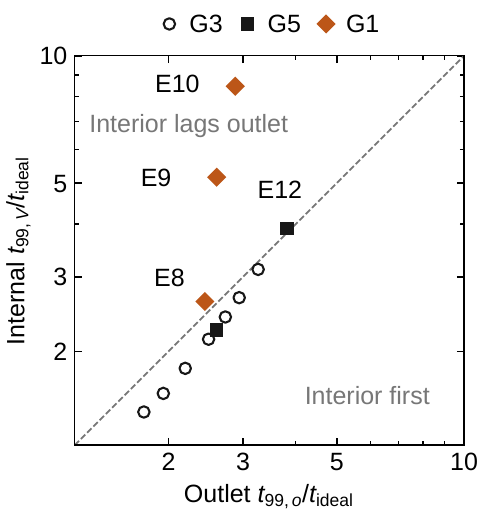}
\caption{\label{fig:parity}The outlet signal can mislead in both
directions. Internal against outlet 99\% replacement time for all twelve
tracer simulations, both scaled by the ideal flush time. Points below the
dashed diagonal are replaced inside before the outlet shows it; points
above lag, most strongly in the G1 pocket case E10. Symbols mark the pore channel (G1, G3, G5); E8, E9 and E10 are runs in G1 and E12 a run in G5 (all runs in
Table~\ref{tab:tracer}).}
\end{figure}
\subsection{Does the outlet show internal replacement?}\label{sec:res_repl}

We test Q2 by comparing the outlet and internal replacement times in each simulation. Figure~\ref{fig:parity} plots the internal against the
outlet 99\% time. Above the diagonal, the outlet reaches the threshold before the interior, so the outlet signal is optimistic. In eight of the twelve
simulations the interior is replaced 4--18\% before the outlet shows it,
because diffusion carries fresh solute across streamlines into the slow
near-wall fluid. In four simulations the interior lags. Two of these lags
are marginal (factors 1.03 and 1.08), but in the G1 pocket case at
$\Pe=138$ the interior needs $8.5\tid$ against $2.9\tid$ at the outlet,
almost three times longer, and at $\Pe=69$ twice as long
(Table~\ref{tab:tracer}).

The fitted exchange number $\Lambda_e$ orders the internal-to-outlet ratio
across all twelve simulations (Spearman coefficient 0.99; both
quantities come from the same internal transient), with the lag
appearing only where $\Lambda_e$ exceeds about one (Fig.~\ref{S-fig:s_limits}(c)).
In G1, a sixfold change in flow has little effect on the exchange time, which varies nearly inversely with $D$
(Fig.~\ref{S-fig:s_exchange}), so exchange with the pocket is mainly
diffusive.
Solute conservation explains the crossover through Eq.~\eqref{eq:balance}. The balance holds in all twelve simulations to within $1.4\times10^{-3}$, and in the
fitted tail the ratio of interior to outlet residual agrees with the fitted $\Lambda_e$ in every run, for example 0.18 in E1 and 5.96 against 5.84
in E10 (Table~\ref{S-tab:balance}).
When both 99\% crossings lie in the common exponential tail,
the interior lags if the tail decays more slowly than one ideal flush time. Because
the scaled problem depends only on $\Pe$, the diffusivity sweep in G1
(Fig.~\ref{S-fig:s_exchange}(b)), made with the in-house scalar solver, is
equivalent to simulations at $\Pe=69$ and 276. That solver reproduces
the Fluent results for E10 at the baseline diffusivity and for
E9 at $\Pe=69$ to within 1\% (Sec.~\ref{S-sec:verification}).
The lag in G1 thus grows steadily with $\Pe$:
1.08 in E8 ($\Pe=22.6$), 1.99 in E9 ($\Pe=69$), 2.94 in E10 ($\Pe=138$) and more than
3.6 at $\Pe=276$, where the interior had not reached 99\% after twelve flush
times. The lag is a late-time effect: at 90\% replacement the
interior leads the outlet in all twelve simulations, including the pocket
cases, and the ordering reverses only when the threshold falls within the
exchange tail (Table~\ref{S-tab:recon}).
\begin{figure}[!htb]
\centering\includegraphics[width=0.8\textwidth]{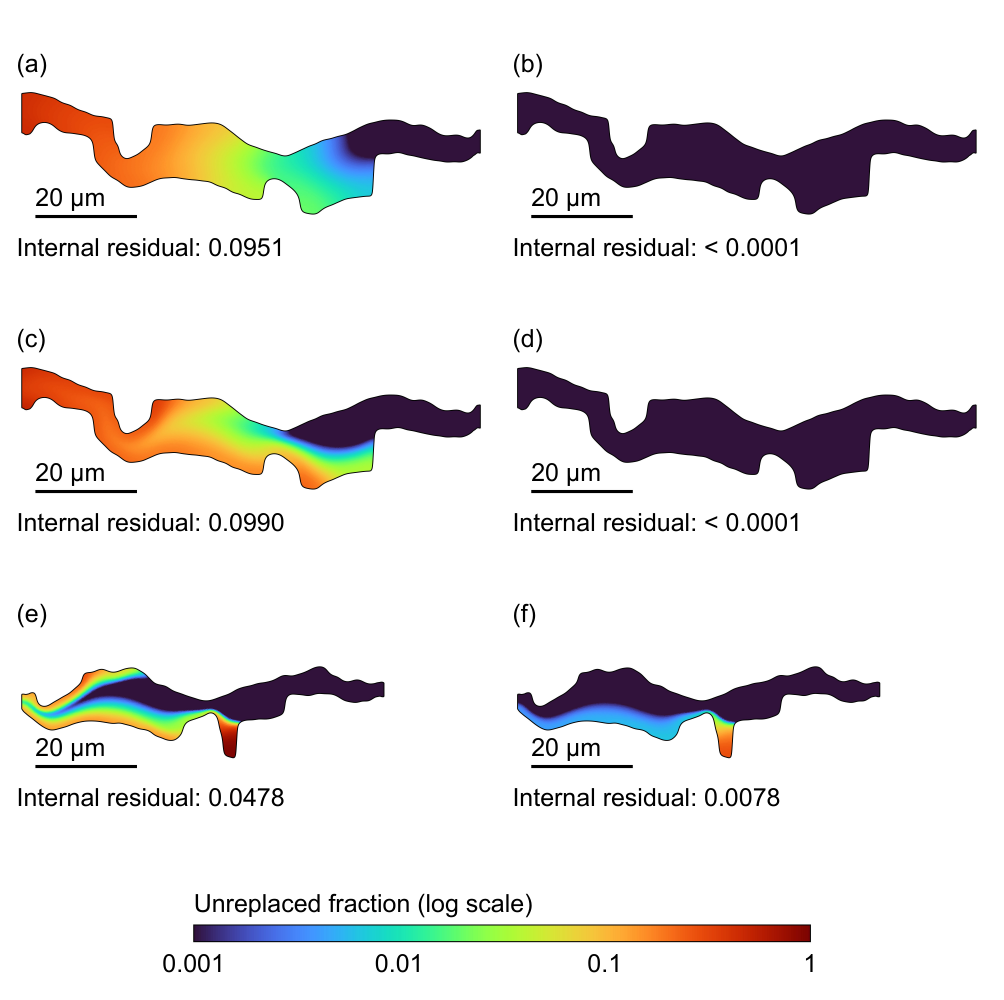}
\caption{\label{fig:snapshots}Old fluid survives in the G1 pocket long
after the main channel is flushed. Unreplaced fraction (logarithmic scale) in the reference channel G3 for
run E1 ($\Pe=3.6$) after (a) one and (b) three flush times and for run E2
($\Pe=21.5$) after (c) one and (d) three flush times, and in channel G1,
which contains the side pocket, for run E10 ($\Pe=138$) after (e) two and
(f) ten flush times. Values below $10^{-3}$ are shown at the bottom of the
color scale; the internal residual printed under each panel is the area
average over the full field.}
\end{figure}

The tracer fields show where the late residual persists (Fig.~\ref{fig:snapshots}). In G3 the
area-averaged residual is about 0.1 after one flush time and falls below
$10^{-4}$ after three,
at both low and moderate $\Pe$. In G1, after two flush times the old fluid
remains along the walls and in the pocket; after ten it is confined to the
pocket, while the main channel has long been refreshed.
The pore simulations show when the outlet can signal replacement before the interior is refreshed. Applying these results to the plug requires comparing the speeds in the tracer runs with those inside the plug.

\subsection{Flow distribution and replacement at the plug scale}\label{sec:res_plug}

\begin{figure}[!htb]
\centering\includegraphics[width=\textwidth]{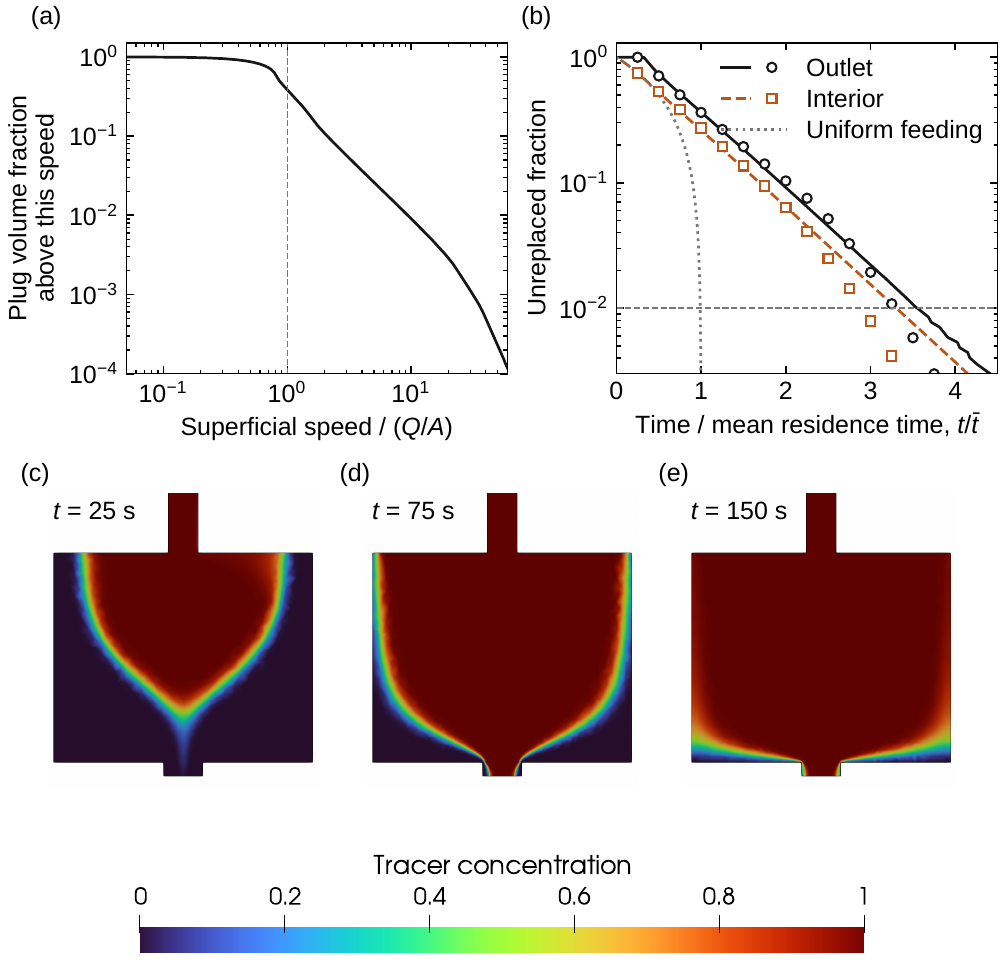}
\caption{\label{fig:plugscale}Port feeding concentrates the flow
near the ports and broadens the residence times of the plug. (a) Fraction
of the plug volume whose superficial speed exceeds a given multiple of
the uniform value $Q/A$ (dashed line). (b) Unreplaced fraction $1-c$ at the
plug outlet port (solid) and in its interior (dashed) in the advective limit,
against time in mean residence times; symbols give the
three-dimensional simulation of the device, which includes
diffusion: the device outlet, shifted by the mean transit times of both
conduits (4.87~s), and the plug average, shifted by that of the inlet
conduit (4.45~s); the dotted curve is the interior of
a uniformly fed plug and the horizontal line marks 99\% replacement.
(c--e) Tracer concentration $c$ on the axial plane of the
three-dimensional device at 5~mL/min, 25, 75 and 150~s after feeding
starts (unshifted time) (about 0.3, 0.9 and 1.9 mean residence times); the last old fluid
lies along the floor of the plug.}
\end{figure}

The two pore velocity mappings represent different regions of the plug.
Figure~\ref{fig:plugscale}(a) shows the distribution of superficial
speed in the port-fed plug. About 38\% of the plug volume moves faster
than the uniform value $Q/A$ and 13\% at less than half of it; only
about 1\% exceeds nine times $Q/A$, and that fluid lies next to the
ports. The Darcy mapping therefore represents the bulk of the plug, with
$\Pe=3.6$--21.5 in G3 over 5--30~mL/min, whereas the conduit mapping
corresponds to about 1\% or less of the plug, next to the ports.

For a G1-like pocket with the fitted G1 exchange time, the local
speed determines whether the exchange number exceeds one. At 5~mL/min
the threshold is about seven times $Q/A$, a speed exceeded in about
1.5\% of the modeled plug. At 30~mL/min the threshold drops to about
1.2 times $Q/A$ and is exceeded in nearly 30\% of the modeled plug.
These multiples are the inverse bulk exchange numbers of G1, with Darcy
flush times of about 200 and 34~ms (Sec.~\ref{S-sec:plug_transport}). A G1-like
pocket in these regions can therefore lag the local outlet at late times.


The uneven flow also broadens the distribution of residence times in the plug (Fig.~\ref{fig:plugscale}(b)). In the advective limit the first
fluid arrives after $0.33\,\bar t$, half of the outlet flux is fresh after
$0.77\,\bar t$, and 99\% replacement takes about $3.6\,\bar t$ at the outlet
and $3.3\,\bar t$ in the interior, against one mean residence time for
uniform feeding; both values change by less than 1\% over three
grids. Feeding through the ports therefore raises the fluid volume
needed to replace the plug about threefold, in addition to raising the
pressure sevenfold.
The three-dimensional simulation, which resolves the ports and
conduits and includes diffusion, gives half-replacement times of
$0.75\,\bar t$ at the device outlet and $0.55\,\bar t$ for the plug
average. The corresponding advective-limit times are $0.77\,\bar t$
at the plug outlet and $0.54\,\bar t$ for the plug average
(Fig.~\ref{fig:plugscale}(b), Sec.~\ref{S-sec:r1t}). The 99\% times in the three-dimensional run are shorter at $3.3\,\bar t$ at the device outlet and $2.9\,\bar t$
for the plug average, against $3.6$ and $3.3\,\bar t$ at the plug outlet
and interior in the advective limit, and both calculations give the
interior crossing first. The run does not isolate the contributions of
physical diffusion, discretization, conduit spreading and the difference
between the simulated and Darcy flow fields to the shorter tail, so its
cause is left open. Even with the unscaled diffusion coefficient, which exceeds
a porosity--tortuosity estimate about fourfold, diffusion spreads solute
only about 0.6~mm per mean residence time, a plug P\'eclet number
$\phi\delta^2/(D\bar t)\approx4\times10^3$, so physical diffusion is small
on the scale of the plug (Sec.~\ref{S-sec:r1t}).
Figure~\ref{fig:plugscale}(c--e) shows the spatial pattern behind the plug-scale response. Fresh fluid spreads from the inlet port along the roof and down the axis. The last old fluid lies along the floor of the plug, far from the outlet port (all recorded times in Fig.~\ref{S-fig:s_r1t}). The
spheroids, injected at the apical center under the inlet port
(Sec.~\ref{sec:experiment}), therefore sit, in this homogeneous model, in
the region refreshed first.
Because the simulation treats the plug as a uniform porous block, these
fields show the plug-scale pattern, not exchange at the pore scale. Port
feeding does not by itself make the outlet misleading: the outlet tail has a decay time of about $0.7\,\bar t$ in the advective limit and
$0.4\,\bar t$ in the three-dimensional simulation, shorter than one mean
residence time, so by Eq.~\eqref{eq:balance} the interior leads slightly.
In the three-dimensional simulation the interior residual is 0.40
times the outlet residual in the tail, as Eq.~\eqref{eq:balance} requires
for that time constant. In both calculations of this homogeneous
plug, the interior reaches 99\% replacement before the outlet; in
the pore simulations the largest lag occurred in the G1 pocket, and G5 run
E12 showed a small lag.


\section{Discussion}
\subsection{Feeding and permeability control the pressure budget}
For Q1, the homogeneous Darcy model separates the effects of feeding and permeability in Eq.~\eqref{eq:capacity}. Comparing
port feeding with uniform feeding at the same $k$ and varying permeability at fixed feeding geometry gives
\begin{equation}
\frac{\Delta p_{\rm port}}{\Delta p_{\rm uniform}}=F,\qquad
\frac{\Delta p(k_1)}{\Delta p(k_2)}=\frac{k_2}{k_1}.
\label{eq:budgetratios}
\end{equation}
The sevenfold feeding penalty is the largest effect in the modeled pressure budget and is set by the boundary conditions independently of pore resistance. For device operation, knowing how the plug is fed and seated in the holder and measuring its permeability are therefore more useful than further refining the pore-scale calculation.

Permeability sets the material contribution to the pressure budget and is its least certain input. Measured
collagen-only matrices at 3 and 6~mg/mL have permeabilities of
$4.7\times10^{-13}$ and $1.2\times10^{-13}$~m$^2$~\cite{avendano2020},
comparable to the G3 value, but at porosities above 0.99 rather than the
assumed 0.27, so the agreement in magnitude does not imply that the pore
structures are alike; collagen microstructure depends strongly on
preparation~\cite{roeder2002,yang2009collagen,raub2007}. Holding the
channel geometry and tortuosity fixed, Eq.~\eqref{eq:permeability} gives
\begin{equation}
\frac{k(\phi)}{k(\phi_0)}=\frac{\phi}{\phi_0},\qquad
\frac{\Delta p(\phi)}{\Delta p(\phi_0)}=\frac{\phi_0}{\phi},
\label{eq:phisensitivity}
\end{equation}
with $\phi_0=0.27$ (Fig.~\ref{S-fig:s_limits}(a)). This brackets the effect
of the assumed porosity but does not replace a measured permeability.

Published permeability measurements put the modeled pressure scale in context. O'Brien \textit{et al.} used a 1.2~m saline head,
approximately 12~kPa, across hydrated collagen--GAG scaffolds~\cite{obrien2007permeability},
while Nasrollahzadeh and Pioletti used a 20~kPa head for synthetic
scaffolds~\cite{nasrollahzadeh2016permeability}. These materials, thicknesses
and supports differ from the present plug, so the comparisons do not
validate its predicted pressure. Varley \textit{et al.} attributed part
of the disagreement between collagen-scaffold permeability measurements
to compression under the imposed pressure gradient and the mounting
arrangement~\cite{varley2016permeability}. Our rigid, sealed-plug calculation specifies the pressure needed to sustain the imposed chamber flow under its assumptions. Testing whether the experimental plug and circuit sustain that state requires measurements of pressure, delivered flow and plug contact.

Compression matters because deformation can change the flow--pressure relation itself. Song \textit{et al.}~\cite{song2019} measured a nonlinear
relation across polydimethylsiloxane membranes and modeled the coupled
changes in membrane dimensions and nanoscale pore structure. A deformable
extension of the present plug model would therefore need to account for
changes in permeability as well as plug dimensions.

Applying the same relations at the reported high pump settings of 40--80~mL/min, which deliver
about 10--20~mL/min to each chamber if the four branches share the flow
equally, gives pressures of about 9--18~kPa with
uniform feeding and 63--127~kPa with port feeding; the delivered flow
at these settings was not measured, and whether the circuit sustains such
pressures, or the plug deforms or detaches first, is unknown. In a dye test at a 5~mL/min setting of the smaller pump, deionized water dyed with methylene blue (1~mg/mL) was
observed downstream of the collagen sponge. The flow through the sponge
and the breakthrough time were not measured, so this setting is not the
5~mL/min plug flow of the calculations, and the test is not compared
quantitatively with the computed residence times.

\subsection{When the outlet misleads}

For Q2, the outlet and interior concentrations weight the pore fluid differently [Eq.~\eqref{eq:advectivelimit}]: the outlet is flux-weighted and the interior is area-weighted. Either ordering is possible in the advective limit. In the computed G3 distribution, near-wall streamlines occupy area and carry little flux, producing a long internal tail. The outlet first sees fresh fluid after $0.61\tid$, while
the last near-wall fluid takes far longer, with an internal 99\% time that
is not converged with wall resolution (Fig.~\ref{S-fig:s_advective}).
Cross-streamline diffusion removes the advective tail, consistent with the interior reaching the threshold first in most of our simulations. In G3 and G5 the ratio
$t_{99,V}/t_{99,o}$ rises from about 0.8 to 1.0 as $\Lambda_\perp$ falls
from 4.6 to 0.02. Within a fixed geometry, however, the definitions of
$\Lambda_\perp$ and $\Pe$ give
\begin{equation}
\Lambda_\perp\Pe=\frac{A_{2D}}{\bar w\,w_{\rm eff}},
\label{eq:mixingpe}
\end{equation}
a purely geometric constant, so $\Lambda_\perp$ is an inverse P\'eclet number
and adds no independent information. The G1 pocket breaks this ordering.

In G1, the competing times are those of flushing the main passage and exchanging fluid with the pocket. Old fluid remains in the pocket after the passage has refreshed when exchange takes longer than one flush. If the residual follows the single exchange mode of
Sec.~\ref{sec:replacement}, $r\simeq B_0e^{-(t-t_0)/\te}$ with $t$ measured
from the inlet switch, and the 1\% crossing lies within that fitted regime
(the fitting window, residuals 0.001--0.03, brackets it), then
\begin{equation}
\frac{t_{99,V}}{\tid}\simeq\frac{t_0}{\tid}+\Lambda_e\ln(100B_0).
\label{eq:tailtime}
\end{equation}
A large exchange number lengthens internal replacement, but the internal-to-outlet ratio also depends on the amplitude $B_0$ and on the
outlet time, which a pocket can itself delay, so Eq.~\eqref{eq:tailtime} is
consistent with the observed association rather than a separate
explanation of it; $B_0$ depends on the time origin and initial loading and
is not simply a stagnant area fraction. The fitted ordering has two further limits. Both
$\Lambda_e$ and $t_{99,V}$ come from the same internal transient, so the
ordering is descriptive rather than an independent prediction; and the
fitted crossover near $\Lambda_e\approx1$ shifts with the
fitting window (Fig.~\ref{S-fig:s_limits}(c)). Predicting $\Lambda_e$ from an
image would require a defined pocket and an independently tested exchange
coefficient. If the matrix may contain dead-end pores, the operator should pair the outlet signal with an internal concentration measurement, for example from sampling or imaging.

The solute balance in Eq.~\eqref{eq:balance} gives a crossover criterion and a way to recover the internal residual from the outlet history. The crossover at $\Lambda_e=1$ does not rest on the four lagging cases alone: it follows from
conservation of solute whenever the threshold is crossed in the
single-mode tail. The balance also expresses the interior residual in terms of the outlet history alone,
\begin{equation}
r(t)=\frac{1}{\tid}\int_t^\infty\bigl[1-\bar c_o(t')\bigr]\,\mathrm dt',
\label{eq:washout}
\end{equation}
so an outlet record that resolves the tail, together with the flush
time, determines the internal replacement. For the twelve simulated outlet histories, we continue the outlet tail exponentially beyond the record. Equation~\eqref{eq:washout} reproduces the directly computed interior
99\% times within 2\% (Table~\ref{S-tab:recon}); without that continuation
the two strongest pocket cases are underestimated by 22\% and 33\%, so
the record must resolve the tail. For a plug, $\tid$ becomes the
mean residence time $\bar t$, and the relation requires a known pore
volume and a sealed plug, since fluid that bypasses the plug reaches the
outlet without replacing the interior. Where these conditions are not
assured, the outlet and internal measurements should be paired.

For charged colloids in dead-end pores, solute gradients can also drive diffusiophoresis and wall diffusioosmosis. Alessio
\textit{et al.}~\cite{alessio2021} modeled how these coupled processes
redistribute particles in a two-dimensional pore. Including these effects
would require an extension of the present passive-solute model and its
no-slip wall condition.

\begin{figure}[!htb]
\centering\includegraphics[width=\textwidth]{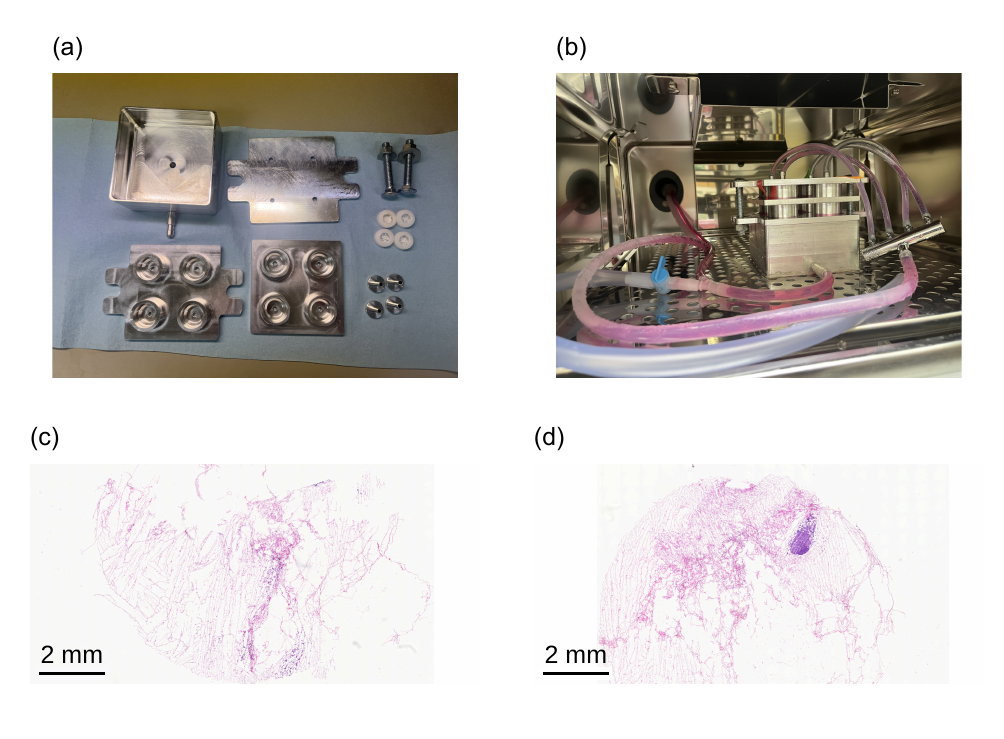}
\caption{\label{fig:experimental}The bioreactor that motivates the
model, and a spheroid-loaded collagen plug. (a) Disassembled device with plug
holders. (b) Assembled circuit with primary and secondary chambers.
(c,d) Hematoxylin-and-eosin sections of a statically incubated plug
two days after spheroid injection, at the first section plane and
100~$\mu$m deeper; scale bars, 2~mm.}
\end{figure}

\subsection{Link to the experiment}\label{sec:experiment}
The experimental bioreactor was developed to support
studies of breast cancer metastasis in a three-dimensional collagen
environment. In such studies, controlled interstitial flow can alter the
direction, mode and dynamics of breast-cancer-cell migration in
three-dimensional matrices~\cite{polacheck2011,haessler2012,polacheck2014,huang2015},
while the same flow carries soluble signals, nutrients and drugs through
the matrix~\cite{swartz2012}. Interpreting these experiments therefore requires knowing the pressure
needed to deliver flow through the matrix (Q1) and whether the outlet signal
reflects replacement of medium within it (Q2).
The simulations use chamber dimensions from this apparatus, which
holds tumor spheroids in collagen. The bioreactor consists of primary and secondary chambers, collagen-plug
holders, a medium reservoir and a removable cover secured with wing nuts
(Fig.~\ref{fig:experimental}(a,b)). A main pump, joined by a second
pump through a Y-connector, feeds a four-way splitter; each branch passes
through a primary and a secondary chamber to the medium reservoir, from
which the medium returns to the main pump through a 0.45~$\mu$m filter.
The reported flow-reversal settings refer to the main pump. The
equal-split estimate assigns each chamber one quarter of the nominal
main-pump flow. The plug is a
commercial collagen sponge (GDT Hemosponge), cut to a cylinder
10~mm in diameter and 10~mm high and soaked for
2~h at 37~$^\circ$C in DMEM containing 1~mg/mL carboxymethylcellulose
(90~kDa); its holder is made to the plug thickness, and the fit
of the hydrated sponge in the holder and its edge seal were not
documented. During perfusion the
medium is high-glucose DMEM with 5\% fetal bovine serum and 1\%
penicillin--streptomycin, kept at 37~$^\circ$C and 5\% CO$_2$ for three
days. Five PANC-1 spheroids, each about 750~$\mu$m in diameter, were
placed in the bevel of an 18-gauge needle with a P1000 pipette and
injected into the apical center of a plug, which was incubated
statically for two days, fixed in 10\% formalin for 12~h, embedded in
paraffin, sectioned at 5~$\mu$m and stained with hematoxylin and eosin at
the NDSU COBRE Histology Facility; the sections were scanned with a
MoticEasyScan Pro~6 with a 20$\times$ objective.
The sections shown in Fig.~\ref{fig:experimental}(c,d) are from the first
section plane and a plane 100~$\mu$m deeper.
The spheroids shown in these sections were prepared from the
pancreatic cancer cell line PANC-1.

The model uses a homogeneous cylinder that fills the chamber, 33.3~mm in diameter and 26.87~mm high, whereas the sponge
was reported cut to 10~mm in diameter and 10~mm high. The modeled plug has the micrometer-scale pores of a dense collagen matrix. The processed sponge sections show larger pore spaces
of order 0.1~mm (Fig.~\ref{fig:experimental}(c,d)), far wider than the
traced channels. Processed sections, however, give neither
hydraulic throat sizes nor a permeability, and the contact of the sponge
with its holder is unknown, so the pressure drop of the experimental
sponge is not established. The value of about 32~kPa is the result for the
modeled chamber-filling plug with the G3 permeability at 5~mL/min
delivered through it, and the experiment provides context for the model
rather than a test of it. The feeding factor and the
replacement criterion do not depend on $k$, but the feeding factor
and the plug-scale residence times depend on the port and plug dimensions
and on how the fluid reaches the plug faces. The values $F\approx7.1$ and
about three pore volumes for 99\% replacement therefore apply to the
modeled chamber-filling plug; the geometry of the holder and its openings
was not documented.

The experiment informs the study in three ways only. The device simulations use the meshed geometry of the chamber
(Fig.~\ref{fig:geometry}(a)) and plug flows of 5--30~mL/min; with the
quarter split these correspond to main-pump settings of about
20--120~mL/min, which include the settings at which flow reversal was
reported. Because the plug obeys Darcy's law, the demand at any other flow
follows by proportion. The plug position and spheroid
placement follow the preparation protocol, and the photographs and
sections document that the device holds a cell-loaded collagen plug.
The operator observations that motivate the two questions
(Sec.~\ref{sec:introduction}) are neither measured chamber flows nor measured reversal
thresholds. No measured pressure, flow or
concentration is used to calibrate or validate the models, and the model
represents one chamber, not the branched circuit, whose branch flows need
not be equal. If the secondary chamber of a branch also holds a plug,
the demand of that branch is the sum of the two.

The circuit uses Kamoer peristaltic pumps (UIP WIFI-S183K,
1--1520~mL/min, and M1-STP, 0--88~mL/min), 1/4-in.\ inner-diameter silicone
and 1/8-in.\ PVC tubing, a Masterflex Y-connector, a stopcock, a four-way
flow-control valve and a 0.45~$\mu$m polyethersulfone capsule filter. The
bioreactor body is milled from aluminum 6061 and the plug holder is 3D
printed in polypropylene.

The present calculations therefore define a conditional pressure budget
and a replacement criterion for this bioreactor, not measured operating
limits. Pressure taps across the plug, branch-flow measurements, a
resolved tracer response and inspection of plug contact would test the
two answers directly and would distinguish among the explanations for the
operator observations.

\subsection{Implications for preclinical cancer models}\label{sec:preclinical}
Perfused three-dimensional tumor models are being
developed for preclinical studies of growth and treatment
response~\cite{marshall2017,goliwas2017}.
Differences in medium delivery between plugs or runs can complicate
interpretation of cell behavior. The pressure results provide a way to assess
the pumping demand for these experiments. At a fixed imposed flow through a
sealed plug, pressure demand scales with the spreading factor and inversely
with permeability, so both the matrix resistance and feeding geometry enter
that assessment.

Medium replacement must also be assessed inside the matrix. Where
cells occupy slowly exchanging regions, dosing or washout schedules based
only on the outlet signal could leave them in contact with old medium for
longer than intended. For a sealed plug of known pore volume and flow rate,
Eq.~\eqref{eq:washout} reconstructs the volume-averaged internal replacement
from an outlet record that resolves the washout tail.
These checks could support reproducible invasion and treatment-response
experiments in perfused collagen models. They are prospective uses of the
framework: the calculations concern a modeled homogeneous plug and a passive
solute and have not been compared with measured cell responses
(Sec.~\ref{sec:limitations}).

\subsection{Model scope and limitations}\label{sec:limitations}
The pore channels sample the effect of segmentation in one healthy aortic image. Biological variation falls outside that sensitivity analysis. The two-dimensional calculations capture viscous resistance through the viscous stresses and no-slip conditions on the resolved pore walls. The geometric reduction does not resolve finite pore depth,
the complete three-dimensional cross-sectional shape or out-of-plane
connections. Additional confinement would increase the resistance of an
otherwise unchanged planar passage, whereas connections outside the
image plane could provide alternative flow paths. The direction and
magnitude of a correction to the inferred permeability therefore cannot
be determined from the traced section alone. Such connections could also
shorten or lengthen pocket exchange in a real plug.

At the plug scale, the three-dimensional calculation represents pore resistance through the prescribed Darcy permeability, with individual pores left unresolved. Agreement with the
axisymmetric calculation checks the feeding factor at that permeability,
but does not independently validate the permeability closure or the
pore-scale replacement mechanism. The predicted pressure demand,
including the estimate at 5~mL/min, therefore remains conditional on
the channel-informed permeability.

The assumptions of rigid walls, homogeneous plug properties and a passive solute exclude plug deformation, matrix remodeling, binding and
cellular uptake, which matter in biological transport~\cite{truskey2009}.
The theory agrees with the simulations near the edge of its formal range,
and the numerical verification supports the reported integral quantities
under the stated settings rather than every local field~\cite{roache1998,asme2009}.

\section{Conclusions}
We asked what sets the pressure needed to perfuse a collagen plug in a
bioreactor, and whether the outlet signal shows that the fluid inside has
been replaced.

For the assumed plug permeability, the pressure is set by the plug, not
by the device, whose pressure difference is only a few pascals.
Inside the plug the resistance arises at narrow, winding throats, and a
second-order theory that follows the channel curvature predicts the
interior pressure drop within about 6\% for the smoother channels, and within
about 13\% when the derivative filter is varied between 0.5 and 1.5~$\mu$m. Among
the modeled factors the feeding geometry matters most: port feeding raises
the demand of the modeled chamber-filling plug about sevenfold over uniform feeding, to about 32~kPa at
5~mL/min through the plug, and a three-dimensional simulation of the full device
confirms this feeding factor for the device ports.

The outlet signal does not reliably report internal replacement. In these
simulations diffusion usually let the interior refresh slightly before the
outlet showed it, but a slowly exchanging pocket made the interior lag
almost threefold, and a fitted exchange number ordered this behavior.
Conservation of solute explains the ordering: when 99\% replacement
is reached in a common exponential tail, the interior lags only where
that tail decays more slowly than one flush time, and the same
balance gives the interior replacement from an outlet record that resolves
the tail. For the modeled plug, port feeding raises the fluid needed for 99\%
replacement to about three pore volumes without, by itself, making the
outlet misleading.

Predicting the pressure demand of a particular plug requires a
measured permeability and a known feeding geometry. For a sealed plug
of known pore volume and flow rate, an outlet record that resolves the
washout tail can recover internal replacement under the conditions of
Eq.~\eqref{eq:washout}. Where those conditions are uncertain, an internal
concentration measurement can complement the outlet signal.

\section*{Supplemental Material}
The Supplemental Material contains the derivation of the curved-channel
theory, the verification data, all theory comparisons, the G1 exchange
evidence, the velocity mappings, the advective limit, the sensitivity to
porosity, wall slope and fitting window, the open-device flow field,
the plug-scale transit-time calculation, the solute-balance and
outlet-reconstruction checks, and the three-dimensional tracer simulation.
\begin{acknowledgments}
AI-assisted tools were used for grammar checking and computational analysis verification; the authors verified the work and assume full responsibility for its content.
\end{acknowledgments}

\section*{Author contributions}
\textbf{MMHA:} study design, geometry preparation, method development,
numerical simulations, data post-processing, data analysis, writing and
revision; \textbf{MY:} writing; \textbf{KVH:} experimental design,
experiments, writing; \textbf{MS:} bioreactor CAD modeling, writing;
\textbf{SI:} 3D milling code development; \textbf{SM:} experimental development, graduate
student supervision, writing; \textbf{SB:} method development, supervision,
funding acquisition, project administration, writing.
\section*{Funding}
This work was supported by the National Institutes of Health (NIH)
Centers of Biomedical Research Excellence (COBRE) Award 2P20GM109024.
This work was also supported by the Advanced Research
Projects Agency for Health (ARPA-H) under Federal Award No.~140D042690009.
Any opinions, findings, and conclusions or recommendations expressed here
are, however, those of the authors and do not necessarily reflect views
of the funding agencies.
\section*{Conflict of interest}
The authors have no conflicts to disclose.

\section*{Data availability}
The data and code that support the findings of this study are openly
available at \url{https://github.com/akash312010/bioreactor-collagen-plug}.
The source micrograph is available in Ref.~\cite{niestrawska2022}.

\bibliography{references}
\end{document}


\title{Supplemental Material for ``Pressure demand and solute replacement in a collagen-plugged bioreactor''}
\author{Mohammad Mehedi Hasan Akash}
\email[Contact author: ]{makash@eng.famu.fsu.edu}
\affiliation{Department of Mechanical and Aerospace Engineering, FAMU--FSU College of Engineering, Tallahassee, Florida, USA}
\affiliation{Department of Mechanical Engineering, South Dakota State University, Brookings, South Dakota, USA}
\author{Mohammad Yeasin}
\affiliation{Department of Mechanical Engineering, South Dakota State University, Brookings, South Dakota, USA}
\author{Karl Van Horsen}
\affiliation{Department of Pharmaceutical Sciences, North Dakota State University, Fargo, North Dakota, USA}
\author{Mithul Sathish}
\affiliation{Department of Pharmaceutical Sciences, North Dakota State University, Fargo, North Dakota, USA}
\author{Susan Immanuel}
\affiliation{Department of Pharmaceutical Sciences, College of Pharmacy and Allied Health Professions, South Dakota State University, Brookings, South Dakota, USA}
\author{Sanku Mallik}
\affiliation{Department of Pharmaceutical Sciences, College of Pharmacy and Allied Health Professions, South Dakota State University, Brookings, South Dakota, USA}
\author{Saikat Basu}
\email[Contact author: ]{Saikat.Basu@sdstate.edu}
\affiliation{Department of Mechanical Engineering, South Dakota State University, Brookings, South Dakota, USA}
\date{25 November 2026}
\maketitle

\section{Derivation of the curved-channel theory}\label{sec:derivation}
\textbf{Coordinates and sign convention.} The reference curve has arc
length $s$, increasing from the outlet ($X=0$) to the inlet ($X=L_c$), unit
left normal $n$, signed curvature $\kappa$ and metric $\mathcal H=1-\kappa n$.
The walls are at $n=c\pm h$, and $q>0$ is the throughflow per unit depth,
directed toward decreasing $s$. With $u_s=-\partial_n\psi$ and
$u_n=\mathcal H^{-1}\partial_s\psi$, continuity holds identically and the
vorticity is $\omega=\mathcal L\psi$, where
\begin{equation}
\mathcal L f=\frac{1}{\mathcal H}
\left\{\partial_s\!\left(\frac{\partial_s f}{\mathcal H}\right)
+\partial_n(\mathcal H\partial_n f)\right\}.
\end{equation}
Stokes flow gives $\partial_s p=-\mu\mathcal H\partial_n\omega$,
$\partial_n p=\mu\mathcal H^{-1}\partial_s\omega$ and hence
$\mathcal L(\mathcal L\psi)=0$, with $\psi=\pm q/2$ and $\partial_n\psi=0$ at
$n=c\pm h$. Local regularity requires $1-\kappa c-|\kappa|h>0$; the bound
$|\kappa|(|c|+h)<1$ is sufficient but not equivalent, and neither proves
global non-overlap of the coordinate map.

\textbf{Expansion.} With the slow coordinate $\chi=\epsilon s$ and
$\kappa=\epsilon K(\chi)$, $h$ and $c$ of order one,
$\mathcal L=\partial_n^2+\epsilon(-K\partial_n)+\epsilon^2(\partial_\chi^2-K^2n\partial_n)+O(\epsilon^3)$.
Writing $\psi=\psi_0+\epsilon\psi_1+\epsilon^2\psi_2$, the flux is carried by
\begin{equation}
\psi_0=\frac q4(3\eta-\eta^3),\qquad \eta=\frac{n-c}{h},
\end{equation}
and the corrections satisfy $\psi_j=\partial_n\psi_j=0$ at both walls and
\begin{align}
\partial_n^4\psi_1&=2K\partial_n^3\psi_0,\nonumber\\
\partial_n^4\psi_2&=2K\partial_n^3\psi_1-2\partial_\chi^2\partial_n^2\psi_0
+K^2\partial_n^2\psi_0+2K^2n\partial_n^3\psi_0,
\end{align}
which give $\psi_1=-(Khq/8)(1-\eta^2)^2$ and a polynomial $\psi_2$
(provided in the data repository). Evaluating
$\partial_s p=-\mu\partial_n\omega$ on $n=0$ gives Eq.~(\ref{M-eq:theory})
of the main text. Two independent symbolic derivations agree term by term,
and no $\mathrm d\kappa/\mathrm ds$ term appears to this order.

\textbf{Checks and evaluation line.} For $c=0$,
$\mathcal C=\tfrac75h_s^2-\tfrac1{10}hh_{ss}+\tfrac2{15}h^2\kappa^2$; a channel
of constant width and curvature gives $w^2\kappa^2/30$. For a flat wall
($\kappa=0$, $c=h$, gap $G=2h$), $\mathcal C=\tfrac35G_s^2+\tfrac1{10}GG_{ss}$,
Eq.~(19) of Tavakol \textit{et al.} normalized by its leading term. Along the
wall-pair midline the pressure gradient $p_s+c_sp_n$ has, at $\kappa=0$, the
correction $\mathcal C_{\rm mid}=2c_s^2+\tfrac75h_s^2-\tfrac1{10}hh_{ss}$; the
two forms differ by an exact derivative. For a straight channel tilted by
$\theta$, the exact relative gradients are $\sec^2\theta$ on $n=0$ and
$\sec^4\theta$ along the midline, whose expansions contain $c_s^2$ and
$2c_s^2$. The integrated forms agree only between parallel-walled ends; the
comparison intervals here are not parallel, so the main-text integral is
evaluated directly on $n=0$.

\textbf{Spreading correlation.} The constriction correlation in
Fig.~\ref{M-fig:capacity}(b) is
\begin{equation}
F_{\rm corr}=1+\frac{\pi R^2}{4\delta}\left[\frac{g(a_1/R)}{a_1}+\frac{g(a_2/R)}{a_2}\right],\qquad g(\xi)=(1-\xi)^{3/2},
\end{equation}
a finite-thickness superposition rather than an exact cylinder solution. It
gives $F=7.04$ for the device ports, 1.2\% below the numerical 7.12 (three
grids: 7.29, 7.09, 7.12).

\section{\texorpdfstring{Axisymmetric Darcy calculation and pressure scaling}{Axisymmetric Darcy calculation and pressure scaling}}\label{sec:darcy_solver}

\textbf{Boundary-value problem.} The in-house finite-volume solver solves
main-text Eq.~\eqref{M-eq:axisymdarcy} in a cylindrical plug, with the
outlet disk at $z=0$ and the inlet disk at $z=\delta$. Writing
$P=(p-p_{\rm out})/(p_{\rm in}-p_{\rm out})$, the port conditions are
$P=0$ for $r<a_2$ on the bottom face and $P=1$ for $r<a_1$ on the top
face. The rest of the two end faces and the cylindrical wall have zero
normal derivative, and the regular axis has $\partial_rP=0$. No gel
deformation, bypass or spatial variation in permeability is included.

\textbf{Finite-volume implementation.} The meridional domain is
divided into cell-centered annular control volumes. For target spacing
$\Delta$, $N_r=\operatorname{round}(R/\Delta)$ and
$N_z=\operatorname{round}(\delta/\Delta)$, with actual spacings
$\Delta r=R/N_r$ and $\Delta z=\delta/N_z$. Integration of
$\nabla\cdot\nabla P=0$ gives
\begin{equation}
\sum_f \frac{A_f}{d_f}(P_N-P_P)=0,
\label{eq:darcy_fv}
\end{equation}
where $P_P$ is the cell value, $P_N$ the adjacent cell value, $d_f$ their
separation and $A_f$ the axisymmetric face area. Radial faces use
$A_f=2\pi r_f\Delta z$ and axial faces use
$A_f=2\pi r_i\Delta r$, with $r_i$ the cell-center radius. On a port
boundary the prescribed value replaces $P_N$ and $d_f=\Delta z/2$;
sealed faces contribute no flux. The zero-area radial face at $r=0$
implements the axis condition. A boundary face belongs to a port when
its cell-center radius lies inside the specified disk. The resulting
sparse linear system is solved with a direct sparse solver.

\textbf{Flux and feeding factor.} The positive inlet conductance
of the normalized problem is evaluated from the same boundary-face
fluxes as the discrete balance,
\begin{equation}
\mathcal G=\sum_{r_i<a_1}\frac{2(1-P_{i,N_z})}{\Delta z}
\,2\pi r_i\Delta r,
\qquad
Q=\frac{k}{\mu}\Delta p\,\mathcal G.
\label{eq:darcy_conductance}
\end{equation}
Thus the computed resistance divided by the uniform-feeding resistance
is
\begin{equation}
F=\frac{\mu/(k\mathcal G)}{\mu\delta/(kA)}
=\frac{A}{\delta\mathcal G}.
\label{eq:darcy_normalization}
\end{equation}
Here $\mathcal G$ has dimensions of length when the geometric inputs use
a consistent length unit; $F$ is dimensionless. The outlet flux is
obtained analogously from $2P_{i,1}/\Delta z$ on $r_i<a_2$, providing
a conservation check on the assembled problem. Target spacings of
0.10, 0.05 and 0.025~mm give $F=7.2868$, 7.0887 and 7.1238 for
$R=16.664$~mm, $\delta=26.873$~mm, $a_1=1.93$~mm and
$a_2=2.4995$~mm, retaining the geometry precision used by the code.
Port footprints change discretely under refinement; the nonmonotonic
sequence is reported as grid sensitivity rather than an observed
convergence order. As a limiting check, feeding across both complete
faces gives $F=1$. The port-size sweep uses equal inlet and outlet
radii, while the device calculation retains their unequal radii.

The three-dimensional Fluent calculation can also provide $F$
by the same resistance normalization. Its role here is to test the
axisymmetric, equipotential-port reduction in the full device at
prescribed permeability. Agreement between the two implementations
checks that reduction and does not determine the permeability of the
experimental gel.

\subsection{\texorpdfstring{Lower-flow predictions from the pressure model}{Lower-flow predictions from the pressure model}}\label{sec:pressure_scaling}

The scaling follows directly from main-text
Eq.~\eqref{M-eq:capacity}. With the geometry and material properties
fixed, the hydraulic resistance $R_h=F\mu\delta/(kA)$ is constant.
Using $F=7.12$, $\mu=9.64\times10^{-4}$~Pa\,s,
$\delta=26.87$~mm, $R=16.66$~mm and the G3 permeability
$k=5.57\times10^{-13}$~m$^2$ gives
$A=8.7197\times10^{-4}$~m$^2$ and
$R_h=3.7973\times10^{11}$~Pa\,s\,m$^{-3}$.
Since 1~mL/min is $10^{-6}/60$~m$^3$/s, the pressure--flow coefficient
is 6.3288~kPa per mL/min. At 5~mL/min this gives 31.64~kPa, rounded
to approximately 32~kPa in the main text. Table~\ref{tab:lowerflow}
evaluates the same relation at lower flow rates; none of its rows
represents an additional CFD run.

\begin{table}[!tbp]

\caption{Lower-flow Darcy-model predictions for the sealed,
port-fed G3 plug, with the geometry, viscosity, permeability and feeding
factor held fixed.}\label{tab:lowerflow}
\begin{ruledtabular}
\begin{tabular}{rr}
$Q$ (mL/min) & $\Delta p_{\rm plug}$ (kPa)\\
0.10 & 0.633\\
0.25 & 1.58\\
0.50 & 3.16\\
1.00 & 6.33\\
2.00 & 12.7\\
5.00 & 31.6\\
\end{tabular}
\end{ruledtabular}
\end{table}

For the same plug dimensions and viscosity, departures from this
reference case satisfy
\begin{equation}
\frac{\Delta p}{\Delta p_0}
=\frac{Q}{Q_0}\frac{F}{F_0}\frac{k_0}{k},
\label{eq:pressure_rescaling}
\end{equation}
where the subscript 0 denotes the reference calculation. Changes in
port or plug geometry require a new feeding factor, rather than keeping
$F_0$ by assumption. Changes in plug dimensions also require updated
$A$ and $\delta$ in main-text Eq.~\eqref{M-eq:capacity}.
The scaling also requires constant Newtonian
viscosity, fixed permeability, a rigid plug and sealed contact. It does
not establish the delivered flow of a pump in the branched circuit or
the pressure at which the gel deforms or detaches. Nor does this
pressure scaling imply an inverse-flow scaling of tracer replacement
times, because diffusion and pocket exchange compete with advection.

\subsection{\texorpdfstring{Plug-scale speed and transit-time distributions}{Plug-scale speed and transit-time distributions}}\label{sec:plug_transport}

The distributions in main-text Fig.~\ref{M-fig:plugscale} use the
face fluxes of the axisymmetric solution. The Stokes stream function is
accumulated from the axis across the horizontal faces of each cell row, so
that $\psi=0$ on the axis and $\psi=Q$ on the side wall at every height.
The superficial speed is evaluated at cell centers and scaled by $Q/A$.
For main-text Eq.~\eqref{M-eq:transit}, $\psi$ is interpolated bilinearly
within each cell and sampled at $8\times8$ points; the sampled volume is
collected in 400 flux bins that are refined toward the wall streamline,
and $T=\phi\,\Delta V/\Delta\psi$ in each bin. Where sampling noise in
the narrow wall-side bins would make $T$ decrease toward the wall, adjacent
bins are pooled with flux weighting, which conserves the plug volume. The
flux-weighted mean of $T$ equals $\bar t$ to round-off. For grid spacings of 0.1, 0.05 and 0.025~mm
the first arrival is 0.326, 0.327 and 0.327~$\bar t$, the interior 99\%
time 3.31, 3.31 and 3.30~$\bar t$, and the outlet 99\% time 3.55, 3.56 and
3.55~$\bar t$. Outlet residuals below about $3\times10^{-3}$ belong to the
stream tubes nearest the wall and are not resolved.
The three-dimensional flow field of the production porous-plug run
is used for a tracer check in Sec.~\ref{sec:r1t}.
The volume fractions with speed above 0.5, 1, 2, 9, 12 and 26 times $Q/A$
are 0.87, 0.38, 0.11, 0.011, 0.0068 and 0.0016. Relative to the Darcy
mapping, the conduit mapping multiplies the pore speed by 8.8 in G1, 11.8
in G3 and 26.4 in G5. With the G1 exchange time of about 28~ms from E8 (E10 gives about
22~ms), the bulk
exchange number of that channel, with Darcy flush times $L_cA\phi/Q$ of about 200~ms at
5~mL/min and 34~ms at 30~mL/min, is 0.14 at 5~mL/min and 0.83 at 30~mL/min,
which gives the speed multiples quoted in the main text.

\section{Numerical settings and pressure definitions}\label{sec:numerics}
\textbf{Time discretization.} In the finite-volume balance of the main
text, $\delta_t$ denotes an implicit backward-difference derivative:
\begin{equation}
\delta_t c^{n+1}=\frac{c^{n+1}-c^{n}}{\Delta t}\ \text{(backward Euler)},\qquad
\delta_t c^{n+1}=\frac{3c^{n+1}-4c^{n}+c^{n-1}}{2\Delta t}\ \text{(BDF2)},
\end{equation}
with all fluxes evaluated at the new time level. All tracer simulations
used Fluent's bounded second-order implicit formulation, a bounded form of the
BDF2 scheme above. The time step was $\Delta t=\tid/2000$ in every
run, from $8.50\times10^{-7}$~s in E12 to $1.28\times10^{-4}$~s in E1.

\textbf{Spatial discretization.} All pore cases use second-order
upwind convection for momentum and the scalar, second-order pressure
interpolation and SIMPLE coupling; the open-device case uses SIMPLEC. Diffusive
fluxes use reconstructed face gradients with non-orthogonal correction.
The scalar diffusivity is entered as $\Gamma=\rho D=6.78\times10^{-7}$
kg\,m$^{-1}$\,s$^{-1}$, which with $\rho=1012$~kg/m$^3$ gives
$D=6.70\times10^{-10}$~m$^2$/s.

\textbf{Device pressure.} The open-device simulations include gravity
($g_y=-9.81$~m/s$^2$) at constant density 1012~kg/m$^3$, so the solved
pressure is the modified pressure of main-text Eq.~(\ref{M-eq:ns}). The
device pressure difference is the area-weighted total-pressure difference
between inlet and outlet; on the 2-million-cell grid at 15~mL/min it is
1.89~Pa, and the static-pressure difference is 1.75~Pa.

\section{Constructed geometry and numerical regularization}
The mathematical metric condition applies to the constructed reference
curve and normal wall intersections. Gaussian filtering of $h,c,\kappa$
at 1~$\mu$m is used to regularize the numerical correction, and changes
the diagnostic at the shortest reference smoothings. Table~\ref{tab:metric}
distinguishes these operations. Both representations are locally regular
for the 3 and 4~$\mu$m references. Local positivity is not a proof of
global injectivity or a truncation-error bound.

\begin{table}[!tbp]
\caption{\label{tab:metric}The reference coordinates are regular at the 3 and
4~$\mu$m smoothings used in the main text. First two columns: percentage
of arc length with nonpositive metric at 2~$\mu$m smoothing, for the
constructed and the filtered geometry. Last two: minimum metric over all
comparison intervals at 3 and 4~$\mu$m.}
\begin{ruledtabular}
\begin{tabular}{lrrrr}
Channel & Constructed (\%) & Filtered (\%) & Min., $3~\mu$m & Min., $4~\mu$m\\
G1 & 4.006 & 3.381 & 0.435 & 0.710\\
G2 & 5.652 & 5.428 & 0.063 & 0.522\\
G3 & 2.446 & 0.936 & 0.350 & 0.594\\
G4 & 2.300 & 0.637 & 0.334 & 0.599\\
G5 & 0.337 & 0.000 & 0.389 & 0.626\\
\end{tabular}
\end{ruledtabular}
\end{table}

\section{Verification and theory data}\label{sec:verification}
\begin{figure}[!tbp]
\centering\includegraphics[width=\textwidth]{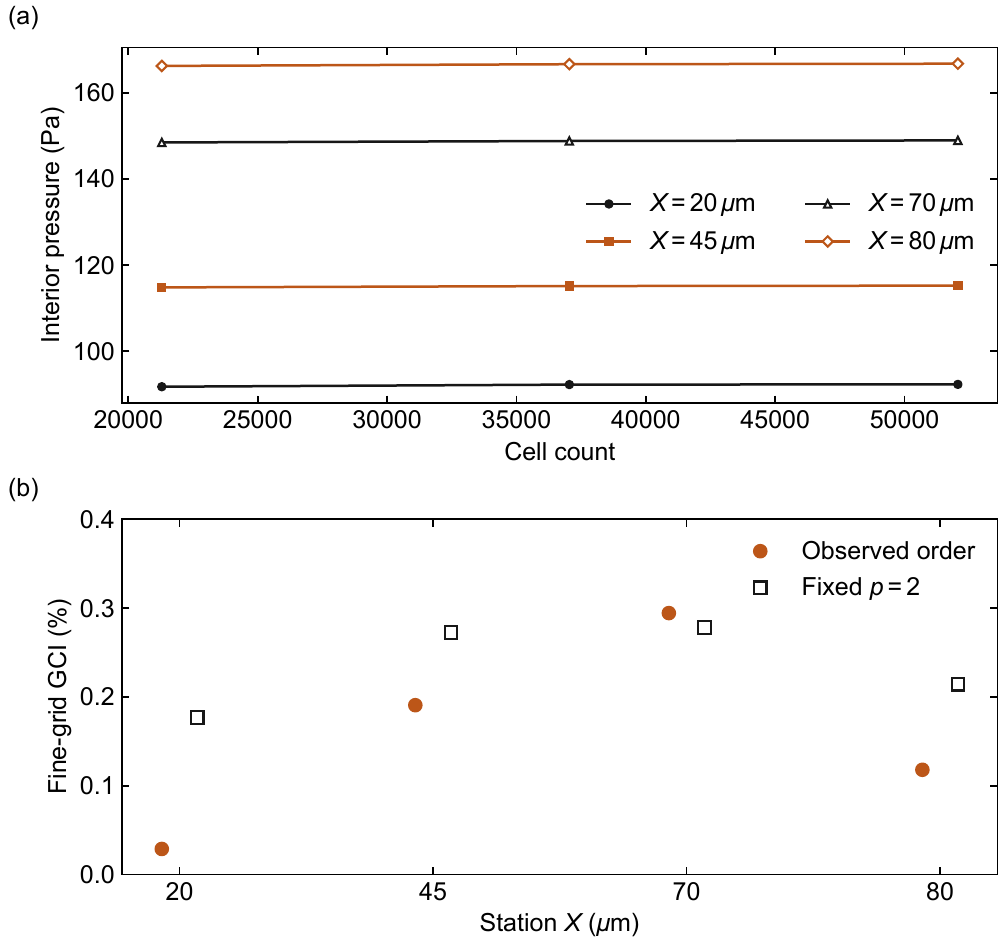}
\caption{\label{fig:s_pressure}The interior pressure of the reference channel G3 is grid-converged to within 0.3\%.
(a) Pressure at four stations on three grids. (b) Fine-grid convergence
index using the observed order and a fixed order of 2. The large apparent
order at $X=20~\mu$m is a local effect.}
\end{figure}

\begin{figure}[!tbp]
\centering\includegraphics[width=\textwidth]{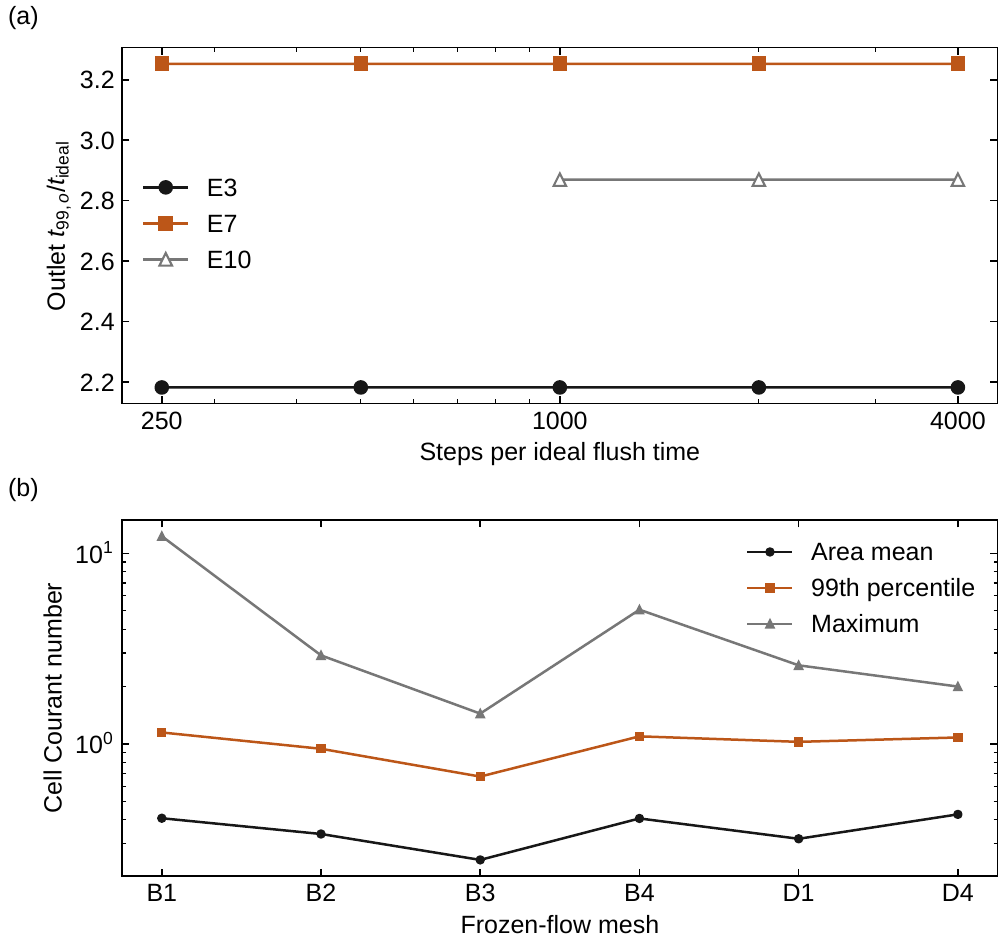}
\caption{\label{fig:s_tracer}Tracer replacement times are insensitive to the time step.
(a) Outlet 99\% time from the independent scalar solver against the
number of time steps per flush time for runs E3 and E7 (G3) and E10 (G1).
(b) Area-mean, 99th-percentile and maximum cell Courant numbers on the
frozen-flow meshes: B1--B3, the fine, medium and coarse G3 grids; B4, the
fine G3 grid with halved first-cell height; D1 and D4, the G1 and G5
grids.}
\end{figure}

\begin{figure}[!tbp]
\centering\includegraphics[width=\textwidth]{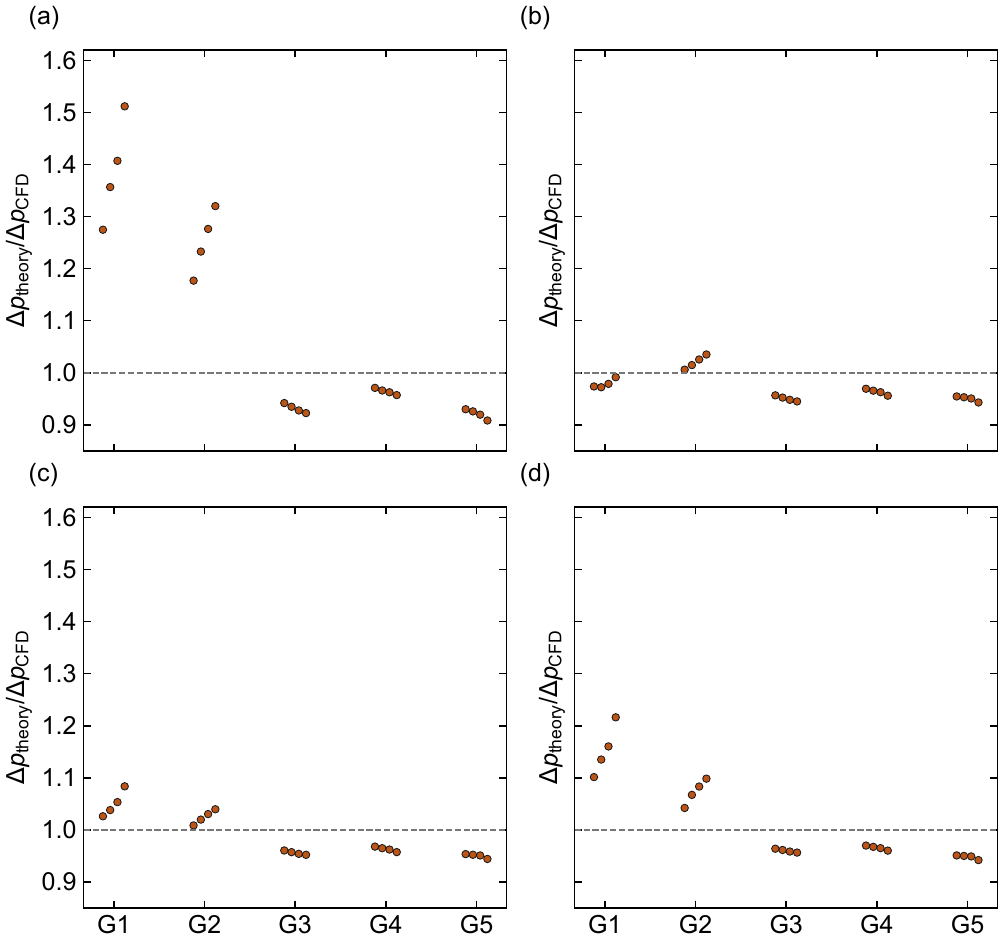}
\caption{\label{fig:s_theory}For G3--G5 the theory stays within about 9\% of the
simulations at every reference smoothing, whereas G1 and G2 are sensitive
to the smoothing (G1--G5 are the five pore channels, Table~\ref{M-tab:geometry}). At the shorter reference smoothings (1 and 2~$\mu$m)
local metric positivity can fail, so these ratios are diagnostic sensitivities
rather than regular-coordinate comparisons (Table~\ref{tab:metric}). Ratio of predicted
to simulated pressure drop for reference smoothings of (a) 1, (b) 2,
(c) 3 and (d) 4~$\mu$m, each with four comparison intervals; the dashed
line marks exact agreement.}
\end{figure}

\begin{figure}[!tbp]
\centering\includegraphics[width=\textwidth]{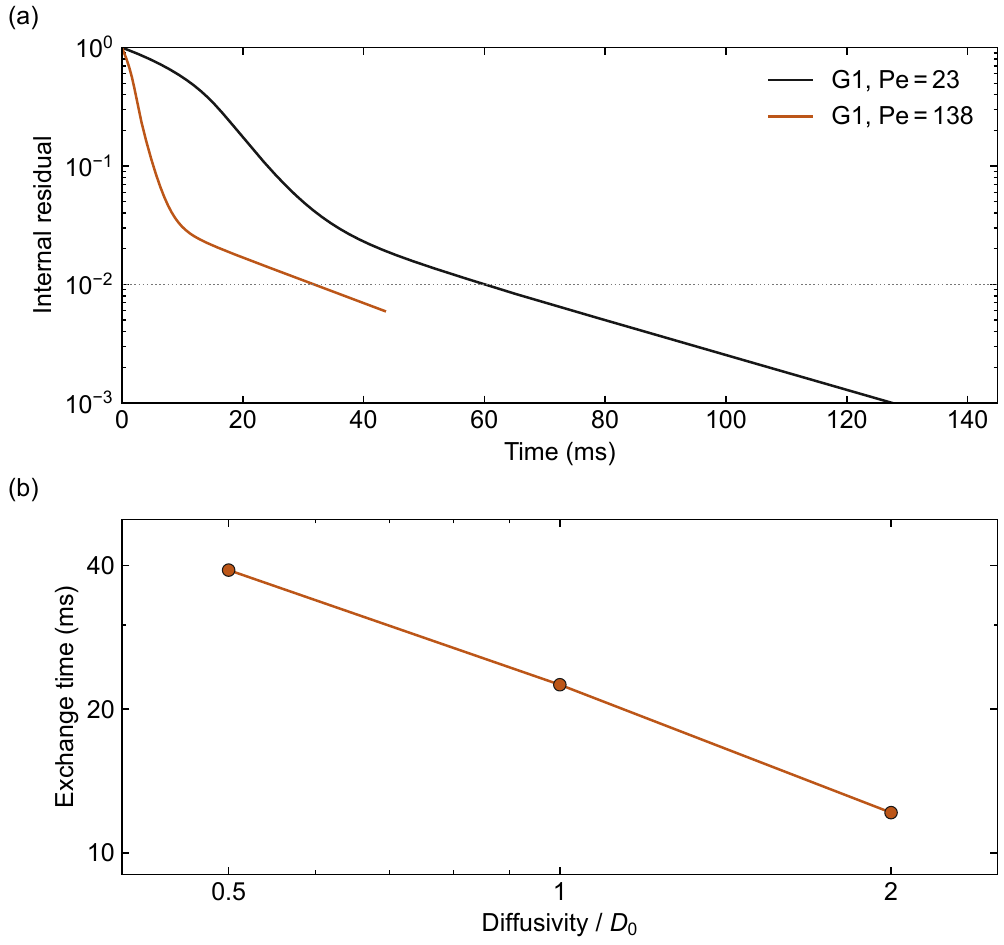}
\caption{\label{fig:s_exchange}Exchange with the G1 pocket is mainly diffusive.
(a) Internal residual against time for runs E8 and E10 in G1, with E10 continued past
its 99\% threshold. (b) Fitted exchange time against diffusivity at fixed
flow; it decreases nearly as $1/D$.}
\end{figure}

\begin{figure}[!tbp]
\centering\includegraphics[width=\textwidth]{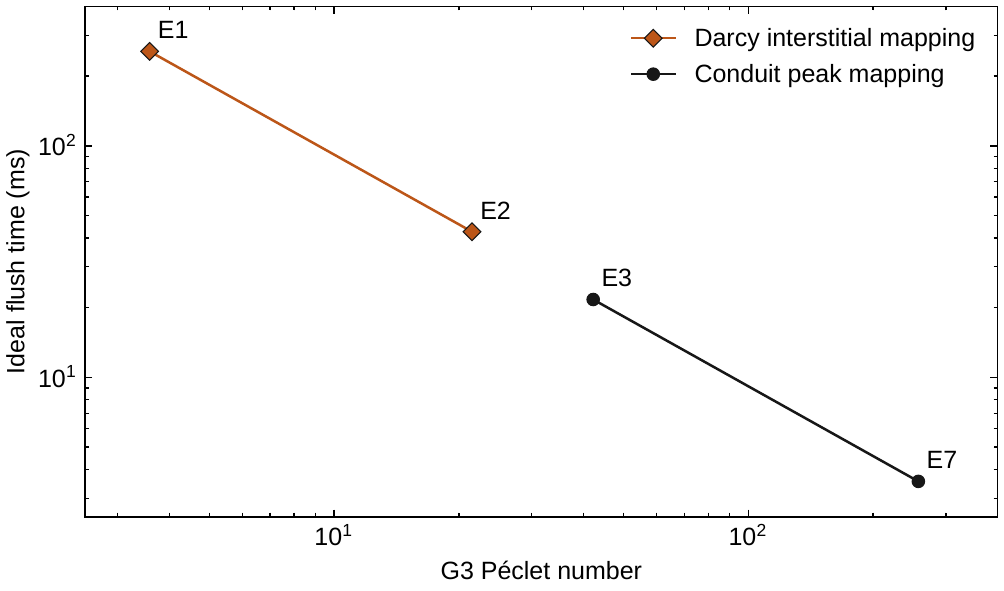}
\caption{\label{fig:s_coupling}The same device flow gives different pore conditions under
the two velocity mappings. Ideal flush time against P\'eclet number for the Darcy
mapping (E1, E2) and the conduit mapping (E3, E7), all in G3, at 5 and
30~mL/min.}
\end{figure}

\begin{figure}[!tbp]
\centering\includegraphics[width=\textwidth]{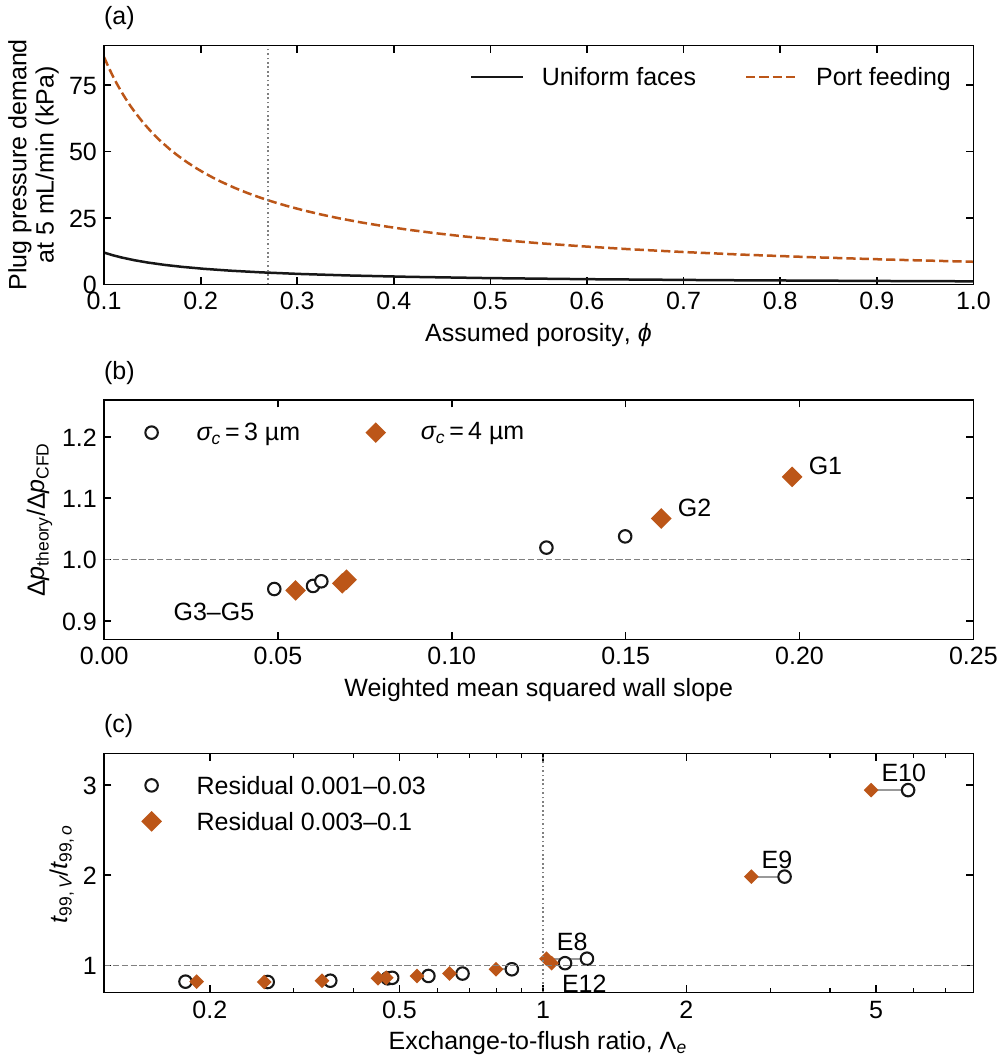}
\caption{\label{fig:s_limits}Sensitivity to the main assumptions.
(a) Plug pressure demand when the porosity in the $k\propto\phi$ closure
is varied at fixed channel geometry ($\phi_0=0.27$), for uniform and port
feeding. (b) Theory-to-CFD ratio against the resistance-weighted wall
slope for the 3 and 4~$\mu$m references; steeper walls give larger
departures. (c) Internal-to-outlet replacement ratio against the exchange
number for two tail-fitting windows (joined by segments); the lag appears
only above $\Lambda_e\approx1$. Labels name the pore channels G1--G5 and the tracer runs E1--E12 listed in Table~\ref{M-tab:tracer} of the main text.}
\end{figure}

Figure~\ref{fig:s_pressure} and Table~\ref{tab:verification} show that the
G3 interior pressure is grid-converged and insensitive to the first-cell
height, and Fig.~\ref{fig:s_tracer} shows that the tracer replacement
times are insensitive to the time step. Figure~\ref{fig:s_theory} and
Table~\ref{tab:theoryranges} give the theory-to-simulation ratio for every
reference smoothing and comparison interval. Figure~\ref{fig:s_exchange}
shows the pocket-exchange tails, Fig.~\ref{fig:s_coupling} the pore
conditions produced by the two velocity mappings, and
Fig.~\ref{fig:s_limits} the sensitivity of the main results to porosity,
wall slope and the tail-fitting window.

\begin{table}[!tbp]
\caption{\label{tab:verification}Grid and wall-layer verification of the G3 interior
pressure on 21\,295, 37\,045 and 52\,059 cells (grids B3, B2, B1; $r_{21}=1.185$,
$r_{32}=1.319$, with $p_o$ and the index computed from the unrounded ratios
$(N_i/N_j)^{1/2}$); the layer check halves the first-cell height from 50 to
25~nm (59\,431 cells, grid B4). The largest index is 0.294\% and the largest layer
change 0.074\%. The high observed order at $X=20~\mu$m reflects local
behavior, not the formal accuracy of the scheme.}
\begin{ruledtabular}
\begin{tabular}{lrrrr}
Quantity & $X=20~\mu$m & $X=45~\mu$m & $X=70~\mu$m & $X=80~\mu$m\\
Coarse (Pa) & 91.735 & 114.797 & 148.450 & 166.220\\
Medium (Pa) & 92.227 & 115.103 & 148.787 & 166.617\\
Fine (Pa) & 92.280 & 115.204 & 148.921 & 166.733\\
Observed order $p_o$ & 7.345 & 2.685 & 1.905 & 3.243\\
Fine-grid GCI (\%) & 0.029 & 0.191 & 0.294 & 0.118\\
Wall-layer change (\%) & 0.074 & 0.021 & 0.049 & 0.040\\
\end{tabular}
\end{ruledtabular}
\end{table}

\begin{figure}[!tbp]
\centering\includegraphics[width=\textwidth]{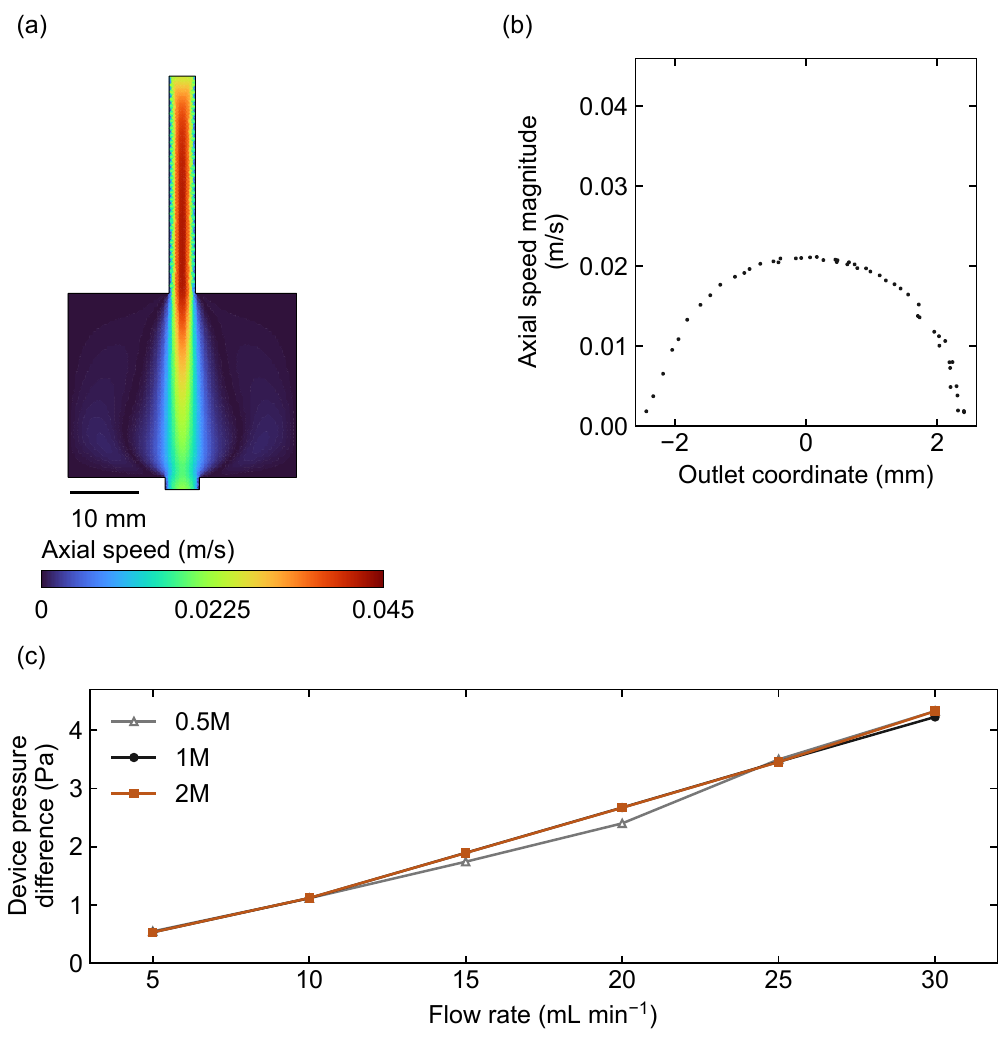}
\caption{\label{fig:s_device}Flow in the open device. (a) Axial speed on
the mid-plane at 15~mL/min (2-million-cell grid). (b) Speed profile in the
outlet. (c) Pressure difference on three grids. With the mass-conserving outlet
mean speed, $\mathrm{Re}_o=22$--134 over 5--30~mL/min.}
\end{figure}

\begin{table}[!tbp]
\caption{\label{tab:theoryranges}Theory-to-simulation pressure-drop ratio over the four
comparison intervals. The first two rows use the 4 and 3~$\mu$m reference curves
with the 1~$\mu$m derivative filter of the main text. The remaining rows
give ranges over both reference curves when the derivative filter is
changed, when the filtered half-width is also used in the leading
$h^{-3}$ term, and when the curvature or offset terms are omitted from
$\mathcal C$.}
\begin{ruledtabular}
\begin{tabular}{lrrrrr}
Case & G1 & G2 & G3 & G4 & G5\\
4~$\mu$m & 1.101--1.216 & 1.042--1.098 & 0.956--0.964 & 0.960--0.970 & 0.942--0.951\\
3~$\mu$m & 1.026--1.083 & 1.008--1.039 & 0.952--0.960 & 0.957--0.968 & 0.944--0.953\\
Filter 0.5~$\mu$m & 1.135--1.350 & 1.077--1.195 & 1.024--1.043 & 1.033--1.050 & 1.018--1.025\\
Filter 1.5~$\mu$m & 0.937--1.073 & 0.948--1.012 & 0.885--0.912 & 0.888--0.921 & 0.873--0.903\\
Filter 2~$\mu$m & 0.873--0.955 & 0.892--0.941 & 0.828--0.876 & 0.832--0.886 & 0.823--0.868\\
Filtered $h$ in $h^{-3}$ & 0.852--0.886 & 0.872--0.913 & 0.890--0.919 & 0.913--0.935 & 0.908--0.929\\
Curvature omitted & 1.008--1.215 & 0.979--1.090 & 0.902--0.922 & 0.909--0.932 & 0.891--0.909\\
Offset omitted & 0.904--1.025 & 0.887--0.947 & 0.742--0.826 & 0.759--0.840 & 0.732--0.811\\
\end{tabular}
\end{ruledtabular}
\end{table}
The agreement of the theory in G3--G5 depends on the derivative filter.
For filter widths of 0.5--1.5~$\mu$m it ranges from 2--5\% above to
8--13\% below the simulations, and with the filtered half-width in both
terms it is 7--11\% low, compared with 20--33\% low at leading order.
The offset terms carry most of the second-order gain, raising the ratio
from 0.73--0.83 to 0.89--0.93, and the curvature terms add about 4--5
percentage points; curvature without offset leaves the ratio close to
its value without either. The ratios are evaluated from the saved
geometry with the simulated pressure drops of the main comparison.
Across the 2--4~$\mu$m set the $h^{-3}$-weighted mean correction in G3--G5 is
0.20--0.33, and its local magnitude exceeds 0.5 on about 17--27\% of the
interval. Omitting curvature and offset together underpredicts by 15--27\%
(17--27\% for the 3 and 4~$\mu$m references of the main text). On the
reference interval (end cuts 6 and 12~$\mu$m) the share of the
leading-order resistance $\int w^{-3}\,\mathrm ds$ is 46--50\%, 40--43\%,
64--65\%, 72--73\% and 80\% for G1 to G5 (2 and 4~$\mu$m reference curves),
and 31--35\% to 63\% for the widest cuts of 10 and 20~$\mu$m.
The open-device grids (about 0.5, 1 and 2 million cells) differ by 0.3--10.2\%
depending on flow (Fig.~\ref{fig:s_device}(c)) and are reported as a
sensitivity rather than a GCI. The
independent scalar solver (central advection, BDF2 or Crank--Nicolson)
reproduces integrated replacement times to about 0.5\%. On the G3
grids B1--B4 (52\,059, 37\,045, 21\,295 and 59\,431 cells, B4 with refined
near-wall layers) at the same time step, its outlet and interior 99\%
times change by at most 0.40\% and 0.49\% in E3 and 0.36\% and 0.59\% in
E7. In E10 the two solvers
give $t_{99,V}/t_{99,o}=2.95$ and 2.94 on the same 35\,351-cell mesh; its central
advection overshoots locally (about 1.22 during the front passage,
with larger start-up extrema of about 1.87), so its agreement in integrated
times does not imply pointwise boundedness of either solver.

\textbf{Pocket channel G1 on a refined mesh.} As a two-mesh sensitivity
check of the G1 pocket lag, the original G1 mesh (D1, 35\,351
quadrilateral cells) was refined by splitting every cell into four through
its edge midpoints and centroid. This gives 141\,404 cells, a refinement
ratio of 2 and half the first-cell height, with the wall outline and zone
definitions unchanged. On the refined mesh the steady flow was solved
again to residuals of about $10^{-8}$ in Fluent 2025 R2, and runs E8 and
E10 were repeated with the original inlet speeds, scalar settings and fit
window. Each run started with a smaller time step ($\tid/19\,275$ for E8
up to $1.6\,\tid$ and $\tid/3160$ for E10 up to $3.1\,\tid$) and continued
with the original $\tid/2000$, so the late-time tail, where the pocket
controls replacement, was computed with the original time step. Because
the early parts used smaller steps, the differences in
Table~\ref{tab:g1refine} describe the two configurations tested, not a
formal estimate of the spatial error. The early replacement
times agree within 0.2\%, and the 99\% times, lag ratios and exchange
times change by at most 2.4\%. In both runs the refined mesh gives a
slightly longer interior time, which may reflect slightly less numerical
diffusion across the pocket mouth; the lag persisted after refinement. The interior-to-outlet residual ratio
in the tail changes by 1.2\% (E8) and 0.4\% (E10), and the solute balance
of main-text Eq.~\eqref{M-eq:balance} holds within $2\times10^{-3}$ on the refined
mesh.

\begin{table}[!tbp]
\caption{\label{tab:g1refine}The G1 pocket lag persisted after mesh
refinement (two-mesh sensitivity check). Replacement times of runs E8 and E10 in flush times $\tid$ and exchange
time $\te$ on the original mesh D1 and on the mesh refined by splitting
every cell into four (refinement ratio 2); the third row of each run gives
the relative change.}
\begin{ruledtabular}
\begin{tabular}{llrrrr}
Run (Pe) & Mesh (cells) & $t_{99,o}/\tid$ & $t_{99,V}/\tid$ & $t_{99,V}/t_{99,o}$ & $\te$ (ms)\\
E8 (22.6) & D1 (35\,351) & 2.44 & 2.62 & 1.08 & 28.3\\
 & refined (141\,404) & 2.46 & 2.69 & 1.09 & 28.8\\
 & change & $+1.0$\% & $+2.4$\% & $+1.4$\% & $+1.9$\%\\
E10 (138) & D1 (35\,351) & 2.88 & 8.47 & 2.94 & 21.9\\
 & refined (141\,404) & 2.88 & 8.60 & 2.99 & 22.3\\
 & change & $+0.1$\% & $+1.5$\% & $+1.4$\% & $+1.6$\%\\
\end{tabular}
\end{ruledtabular}
\end{table}

\textbf{Tracer details.} The Schmidt number is about $1.4\times10^3$ and the
channel Reynolds number 0.0025--0.51. Tail constants are fitted over
residuals 0.001--0.03, with 0.003--0.1 as a sensitivity window, which moves
$\Lambda_e$ from 1.24 to 1.02 (E8), 1.11 to 1.04 (E12) and 5.84 to 4.88 (E10)
and the Spearman coefficient from 0.991 to 0.982. The G1 exchange time
changes from about 28 to 22~ms over a sixfold flow change; at fixed flow it
is 40.2, 22.5 and 12.1~ms for $D/D_0=0.5$, 1 and 2 (exponent about
$-0.86$). For a single-mode tail $t_{99,V}\simeq\te\ln(B/0.01)$, where the
amplitude $B$ depends on the time origin and is not a stagnant area fraction.

\textbf{Image dark-area fraction.} The threshold intensities of
main-text Table~\ref{M-tab:geometry} were set as percentiles within the
traced region of the second-harmonic-generation image. Over the full image,
the fraction of pixels at or below these intensities is 0.36, 0.38, 0.44,
0.49 and 0.53 for G1 to G5.
These fractions are the upper estimates of the pore space used for the
porosity sensitivity in the main text; with $k\propto\phi$ they would
divide the plug pressure demand by 1.3, 1.4, 1.6, 1.8 and 1.9.
\textbf{Pore-volume convention.} Main-text Eq.~\eqref{M-eq:permeability}
assigns each channel the pore volume of a uniform slot of width
$w_{\rm eff}$. With the traced area $A_{2D}$ instead, $k$ is multiplied by
$Sw_{\rm eff}/A_{2D}=0.71$, 0.71, 0.78, 0.81 and 0.84 for G1 to G5 (computed
from $S=\mathcal T L_c$, $w_{\rm eff}$ and $A_{2D}$ in main-text Table~\ref{M-tab:geometry}), and the plug pressure demand by
1.4, 1.4, 1.3, 1.2 and 1.2.

\textbf{Solute balance.} Table~\ref{tab:balance} tests main-text
Eq.~\eqref{M-eq:balance} on the recorded outlet and internal histories of
the twelve tracer simulations. The largest difference between
$\tid\,\mathrm d\bar c_V/\mathrm dt$ and $1-\bar c_o$ after half a flush
time is below $1.4\times10^{-3}$ in every run. Within the fitting window
the ratio of interior to outlet residual agrees with the fitted
$\Lambda_e$. The scaled tracer problem depends only on $\Pe$, so the G1
simulations at $D/D_0=2$ and 0.5 with the independent scalar solver
correspond to $\Pe=69$ and 276 at the baseline diffusivity. At the
baseline diffusivity the same solver gives outlet and internal 99\% times
of 2.87 and 8.47~$\tid$, a ratio of 2.95 against 2.94 in E10. The two
other diffusivities give
outlet 99\% times of 2.59 and 3.26~$\tid$; the internal 99\% time is
5.13~$\tid$ at $\Pe=69$ and was not reached within 12~$\tid$ at
$\Pe=276$. The Fluent simulation E9 at $\Pe=69$ gives 2.60 and
5.16~$\tid$.

\begin{table}[!tbp]

\caption{Check of the solute balance in the tracer simulations:
fitted exchange number and median ratio of interior to outlet residual
in the fitting window (interior residual 0.001--0.03).}\label{tab:balance}
\begin{ruledtabular}
\begin{tabular}{llrrr}
Run & Ch. & $\Lambda_e$ (fit) & $r/(1-\bar c_o)$ & $t_{99,V}/t_{99,o}$\\
E1 & G3 & 0.18 & 0.18 & 0.82\\
E2 & G3 & 0.26 & 0.26 & 0.82\\
E3 & G3 & 0.36 & 0.35 & 0.83\\
E4 & G3 & 0.47 & 0.47 & 0.86\\
E5 & G3 & 0.57 & 0.58 & 0.88\\
E6 & G3 & 0.68 & 0.69 & 0.91\\
E7 & G3 & 0.86 & 0.87 & 0.96\\
E8 & G1 & 1.24 & 1.22 & 1.08\\
E9 & G1 & 3.21 & 3.23 & 1.99\\
E10 & G1 & 5.84 & 5.96 & 2.94\\
E11 & G5 & 0.48 & 0.48 & 0.86\\
E12 & G5 & 1.11 & 1.13 & 1.03\\
\end{tabular}
\end{ruledtabular}
\end{table}

\textbf{Reconstruction from the outlet history.}
Table~\ref{tab:recon} applies main-text Eq.~\eqref{M-eq:washout} to the
recorded outlet histories. The integral over the record is evaluated by
the trapezoidal rule, and beyond its end the outlet residual is continued
as a single exponential fitted to its last factor-of-three decrease. The
reconstructed interior 99\% times agree with the direct ones within 2\%.
The continuation matters only where the record ends with an outlet
residual near $10^{-3}$: without it, E9 and E10 are underestimated by 22\%
and 33\%. The last three columns give the ratio of interior to outlet
replacement time at 90\%, 95\% and 99\%. The interior leads at 90\% in
every run, and the pocket runs reverse only at the late threshold.

\begin{table}[!tbp]

\caption{Interior replacement reconstructed from the outlet history
alone, and the ordering of interior and outlet at three thresholds.
Interior 99\% times in flush times $\tid$.}\label{tab:recon}
\begin{ruledtabular}
\begin{tabular}{llrrrrr}
 & & \multicolumn{2}{c}{$t_{99,V}/\tid$} & \multicolumn{3}{c}{$t_{\alpha,V}/t_{\alpha,o}$}\\
Run & Ch. & Direct & From outlet & 90\% & 95\% & 99\%\\
E1 & G3 & 1.44 & 1.44 & 0.75 & 0.78 & 0.82\\
E2 & G3 & 1.59 & 1.60 & 0.74 & 0.77 & 0.82\\
E3 & G3 & 1.82 & 1.83 & 0.75 & 0.78 & 0.83\\
E4 & G3 & 2.14 & 2.14 & 0.76 & 0.80 & 0.86\\
E5 & G3 & 2.41 & 2.41 & 0.78 & 0.83 & 0.88\\
E6 & G3 & 2.68 & 2.68 & 0.80 & 0.85 & 0.91\\
E7 & G3 & 3.12 & 3.12 & 0.84 & 0.89 & 0.96\\
E8 & G1 & 2.62 & 2.67 & 0.78 & 0.84 & 1.08\\
E9 & G1 & 5.16 & 5.26 & 0.82 & 0.92 & 1.99\\
E10 & G1 & 8.47 & 8.45 & 0.88 & 1.01 & 2.94\\
E11 & G5 & 2.24 & 2.25 & 0.77 & 0.81 & 0.86\\
E12 & G5 & 3.91 & 3.91 & 0.95 & 0.97 & 1.03\\
\end{tabular}
\end{ruledtabular}
\end{table}


\clearpage
\section{Advective limit and transverse mixing}
\begin{figure}[!tbp]
\centering\includegraphics[width=\textwidth]{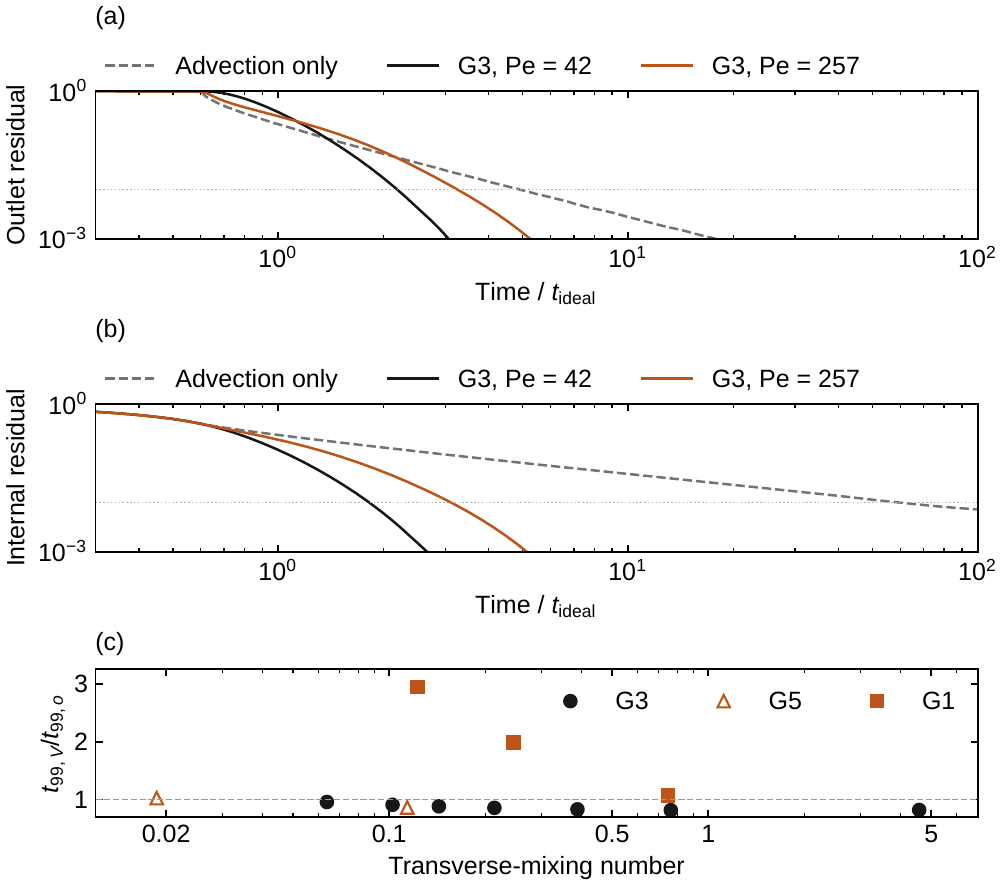}
\caption{\label{fig:s_advective}Diffusion removes the long internal tail of the advective
limit, but not the lag caused by a pocket. (a) Outlet and (b) internal
residual in G3 without diffusion (streamline integration) and with
diffusion (runs E3 and E7). (c) Internal-to-outlet ratio against
$\Lambda_\perp$ for all twelve runs; G3 and G5 follow one trend, while G1
departs from it.}
\end{figure}

Figure~\ref{fig:s_advective} compares the G3 replacement without diffusion,
from streamline integration, with the finite-diffusion simulations. The
streamline calculation uses the same final G3 field as the main
flow figure (52\,059 cells). Iso-streamfunction contours define through-flow
streamtubes; the transit time is the sum of contour-segment length divided
by the speed of the cell containing each segment. Using 2000, 4000, and
8000 contour levels clustered near both walls gives stable first arrival
and outlet quantiles. A separate local repeat with 4000 levels gives
$t_{\rm first}/t_{\rm ideal}=0.6050$ and outlet $t_{50}$, $t_{90}$,
and $t_{99}$ of 0.7018, 1.4475, and 4.9404. The internal 90\% time is
approximately $2.81t_{\rm ideal}$. The residual through-flow area estimate
leaves about 0.6\% in closed recirculation. This last estimate depends on
streamline integration and does not provide a converged internal 99\%
time; that tail is sensitive to wall resolution.

For the actual finite-diffusion simulations,
$\Lambda_\perp=D t_{\rm ideal}/w_{\rm eff}^2$ spans approximately
0.019--4.58. Its product with the reported Pe is constant within each
geometry, so it is not an additional independent parameter. G3 and G5
follow the transverse-mixing trend, whereas G1's deep pocket produces a
large departure. No quantitative developed-dispersion limit is inferred.

\section{\texorpdfstring{Three-dimensional tracer simulation}{Three-dimensional tracer simulation}}\label{sec:r1t}

\begin{figure}[!tbp]
\centering\includegraphics[width=\textwidth]{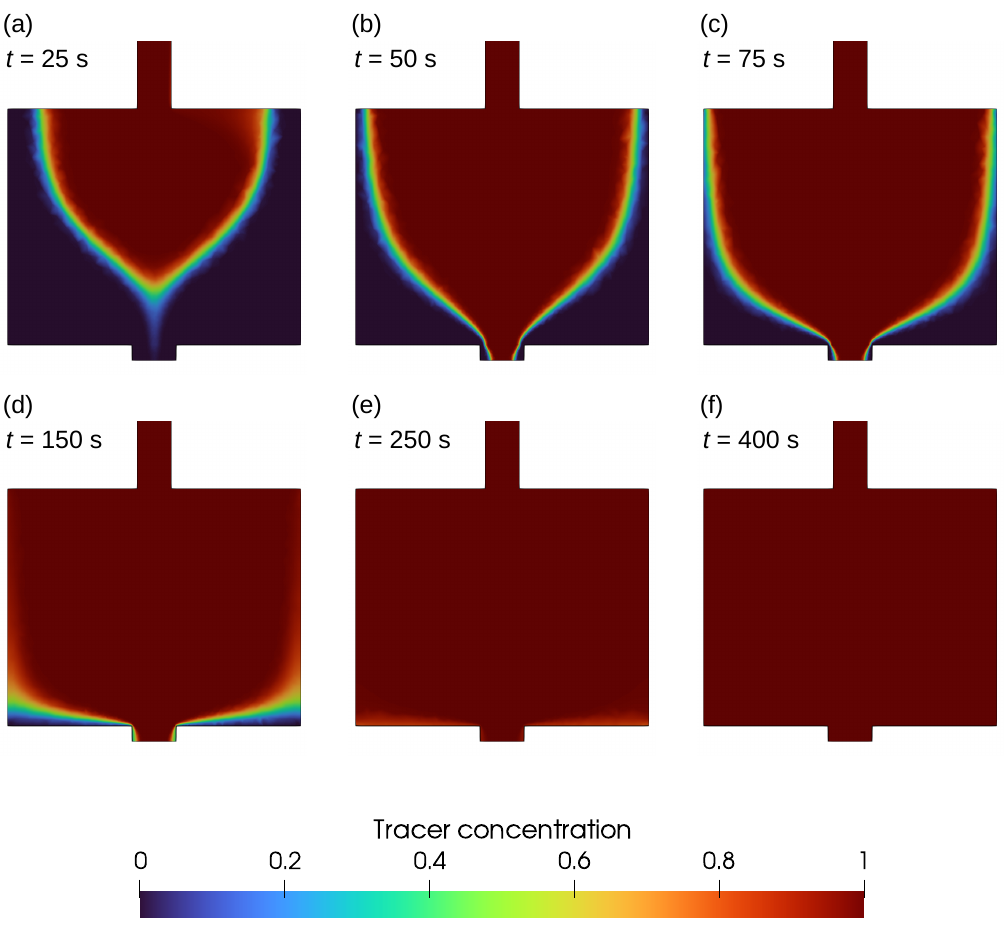}
\caption{\label{fig:s_r1t}Fresh fluid fills the port-fed plug from
the inlet port, and old fluid survives longest along its floor. Tracer
concentration on the axial plane of the three-dimensional device at 5~mL/min
after (a) 25, (b) 50, (c) 75, (d) 150, (e) 250 and (f) 400~s, measured from
the start of feeding; the mean residence time of the plug is 76~s. The
plane shows the plug and the lowest 7~mm of the inlet conduit.}
\end{figure}

The tracer of main-text
Eq.~\eqref{M-eq:tracer} was solved in Fluent on the frozen flow of the
production porous-plug run (4.99~mL/min), with diffusivity
$6.7\times10^{-10}$~m$^2$/s in all zones, porosity 0.27 in the storage term
of the porous zone, advection by the superficial velocity, and the
same coefficient $\Gamma=\rho D=6.78\times10^{-7}$~kg\,m$^{-1}$\,s$^{-1}$ on the
diffusive flux in all zones, with no porosity factor; divided by the
porosity, the plug equation therefore has an effective diffusivity
$D/\phi\approx2.5\times10^{-9}$~m$^2$/s, unit concentration at the inlet from $t=0$, zero
diffusive flux at the outlet and walls, second-order upwind convection,
bounded second-order implicit time integration and a time step of
0.05~s for 420~s. The flow-weighted outlet concentration and the plug
average were recorded at every step. The plug average is shifted by
the mean transit time of the inlet conduit, 4.45~s, and the outlet history
by those of both conduits, $4.45+0.42=4.87$~s; the shift removes the
conduit means but leaves their laminar spreading in the records, and
records at the two ports would be needed for a direct port-to-port
comparison. With these shifts, the integral of the
outlet residual gives a plug residence time of 75.6~s against
$\phi\pi R^2\delta/Q=76.2$~s, a difference of 0.7\%, which confirms that
the porosity enters the storage term. Referred to the plug in this way,
the half-replacement times are $0.75\,\bar t$ at the outlet and
$0.55\,\bar t$ in the interior (0.77 and 0.54 in the advective limit), the
90\% times $2.03$ and $1.71\,\bar t$ (1.94 and 1.69) and the 99\% times
$3.28$ and $2.90\,\bar t$ (3.55 and 3.30); the outlet concentration first
exceeds $10^{-3}$ at $0.27\,\bar t$ (first arrival $0.33\,\bar t$). Between
residuals of 0.03 and 0.001 both histories decay with a time constant of
$0.40\,\bar t$, against $0.70\,\bar t$ in the advective limit, and the
interior residual is 0.40 times the outlet residual, as main-text
Eq.~\eqref{M-eq:balance} requires. Both calculations give the
interior crossing first. The cause of the shorter tail is left open:
physical diffusion, discretization, conduit spreading and the difference
between the simulated and the Darcy flow fields cannot
be separated on this mesh, the only one used for the tracer. Because the
porous-zone flux uses $D$ rather than a porosity--tortuosity value of
order $\phi D/\mathcal T^2$, $D/\phi$ is an upper bound for molecular diffusion
in the pores; even so, the plug P\'eclet number
$\phi\delta^2/(D\bar t)\approx3.8\times10^3$ corresponds to a diffusion length
$(2D\bar t/\phi)^{1/2}\approx0.6$~mm per mean residence time, small
compared with the plug. The inlet conduit carries laminar flow (Reynolds
number about 29, entrance length about 6~mm of its 31.7~mm), so the
centerline tracer arrives about half a conduit transit time early; this
pre-arrival, about $0.03\,\bar t$ for both conduits, gives a plausible scale for part of the difference between the first outlet exceedance at $0.27\,\bar t$ and
the advective first arrival at $0.33\,\bar t$, although the saved records do
not show how much of that offset it explains. Figure~\ref{fig:s_r1t} shows the tracer on the
axial plane at six times, three of which appear in main-text
Fig.~\ref{M-fig:plugscale}(c--e): fresh fluid spreads from the inlet port along the roof and
down the axis, and the last old fluid lies along the floor of the plug,
far from the outlet port.

